\documentclass{emulateapj}

\makeatletter
\usepackage{amsmath} 
\newcommand{\angstrom}{\text{\normalfont\AA}}

\makeatother
\shorttitle{$z\sim3$ Lyman Break Galaxies}
\shortauthors{Bian et al.}

\begin{document}


\title{THE LBT BO\"OTES FIELD SURVEY: I. THE REST-FRAME ULTRA-VIOLET 
AND NEAR-INFRARED LUMINOSITY FUNCTIONS
AND CLUSTERING OF BRIGHT LYMAN BREAK GALAXIES at $Z\sim3$\altaffilmark{1}}


\author{Fuyan Bian, Xiaohui Fan, Linhua Jiang, Ian McGreer}
\affil{Steward Observatory, University of Arizona, 933 North Cherry Avenue, Tucson, AZ 85721, USA}
\author{Arjun Dey}
\affil{National Optical Astronomy Observatory, 950 North Cherry Avenue, Tucson, AZ 85719, USA}
\author{Richard F. Green}
\affil{Large Binocular Telescope Observatory and Steward Observatory, University of Arizona, 933 North Cherry Avenue, Tucson, AZ 85721, USA}
\author{Roberto Maiolino}
\affil{Cavendish Laboratory, University of Cambridge, 19 J. J. Thomson Ave., Cambridge CB3 0HE, UK}

\author{Fabian Walter}
\affil{Max-Planck-Institut f\"ur Astronomie, K\"onigstuhl 17, D-69117 Heidelberg, Germany}
\author{Kyoung-Soo Lee}
\affil{Department of Physics, Purdue University, West Lafayette, IN 47906, USA }
\author{Romeel Dav\'e}
\affil{Steward Observatory, University of Arizona, 933 North Cherry Avenue, Tucson, AZ 85721, USA}

\altaffiltext{1}{Based on data acquired using the Large Binocular Telescope
(LBT). The LBT is an international collaboration among institutions
in the United States, Italy and Germany. LBT Corporation
partners are: The University of Arizona on behalf of the Arizona
university system; Istituto Nazionale di Astrofisica, Italy; LBT
Beteiligungsgesellschaft, Germany, representing the Max-Planck
Society, the Astrophysical Institute Potsdam, and Heidelberg
University; The Ohio State University, and The Research Corporation,
on behalf of The University of Notre Dame, University of
Minnesota and University of Virginia.}


\begin{abstract}
We present a deep LBT/LBC $U_{\rm spec}$-band imaging survey (9 deg$^2$) covering the
NOAO Bo\"otes field. A total of 14,485 Lyman Break Galaxies (LBGs) at $z\sim3$ 
are selected, which are used to measure the rest-frame UV luminosity function
(LF). The large sample size and survey area reduce the LF uncertainties due to Poisson
statistics and cosmic variance by $\ge3$ compared to previous studies.
At the bright end, the LF shows 
excess power compared to the best-fit Schechter function, which 
can be attributed to the contribution of $z\sim3$ quasars. 
We compute the rest-frame near-infrared LF and stellar mass function (SMF) of $z\sim3$ LBGs
based on the $R$-band and [4.5$\mu$m]-band flux relation.
We investigate the evolution of the UV LFs and SMFs between $z\sim7$ and $z\sim3$, 
which supports a rising star formation history in the LBGs.
We study the spatial correlation function of two bright LBG samples 
and estimate their average host halo mass.
We find a tight relation between the host halo mass and the galaxy star formation rate (SFR),
which follows the trend predicted by the baryonic accretion rate onto the halo, suggesting 
that the star formation in LBGs is fueled by baryonic accretion through the cosmic web.
By comparing the SFRs with the total baryonic accretion rates,
we find that cosmic star formation efficiency 
is about 5\%-20\%  and it does not evolve significantly with redshift, halo mass,
or galaxy luminosity.
\end{abstract}


\keywords{galaxies:evolution -- galaxies:formation -- galaxies:high-redshift }



\section{INTRODUCTION}

The redshifts between $1<z<3$ were the most active epochs of galaxy formation, 
when the star formation rate (SFR) density and the activity of bright quasars 
reached their peaks \citep[e.g.,][]{Madau:1996uq, Fan:2001lr}.
In this epoch, the Hubble sequence observed in the nearby Universe
was being built up and about 50$\%$ of the present-day stars
 formed \citep{Dickinson:2003fk}. Therefore,
observations in this redshift range provide crucial clues
to understanding the formation and evolution of galaxies.
The Lyman break technique has been well developed for surveying galaxies in this redshift range
\citep[e.g.,][]{Steidel:1996fk,Steidel:1996lr}.
Large samples of high-redshift star-forming galaxies have
been established with this method \citep[e.g.,][]{Steidel:2003kx}.

Using samples of Lyman break galaxies (LBGs), the rest-frame UV luminosity functions (LFs) from $z\sim2$ to 
$z\sim7-8$ have been well studied. 
\citep[e.g.,][]{Ly:2009qy,Reddy:2009kx,Steidel:1999lr,Sawicki:2006fj,Bouwens:2007lr,Bouwens:2008fk,Yan:2011qy}.
The rest-frame UV LF is a fundamental tracer of  galaxy formation and evolution; it is used to compute the UV luminosity density by applying a dust extinction correction
and to constrain the history of star formation \citep[e.g.,][]{Madau:1996uq}. 
However, LF measurement remains uncertain for 
$z\sim2-3$ LBGs, the measured faint end slope ($\alpha$) of the Schechter function
 ranges from the shallowest with $\alpha=-1.05$ to the steepest with
$\alpha=-1.88$ \citep[e.g.,][]{Ly:2009qy,Reddy:2008fj};
at the bright end, 
there are discrepancies between \citet{Steidel:1999lr}
and \citet{Le-Fevre:2005ys} for the galaxies at $z\sim3$ by factors $1.6-6.2$, and $z\sim4$
by factors $2-3.5$.  In addition, the evolution of the bright end UV LF of 
high-redshift LBGs is still not well constrained. 
\citet{Sawicki:2006lr} claimed that the number density of bright LBGs decreases with redshifts, 
while \citet{Bouwens:2007lr} found that the number density remains constant. 
Furthermore, most of the small area surveys lack  
information on the most luminous LBGs, i.e., with $M_{1700\rm\angstrom} < -23$, 
due to the small surface density of these luminous LBGs.

Galaxy clustering can be used to test the hierarchical theory of structure formation, 
which predicts that the clustering of dark
matter halos strongly depends on their masses and assembly history
\citep[e.g.,][]{Mo:1996kx}. 
Numerical simulations 
can predict the dark matter distribution given the underlying cosmology and 
the initial matter power
spectrum derived from the cosmic microwave background measurements
\citep[e.g.,][]{Spergel:2007fk}. Distributions of galaxies and 
dark matter are connected by the halo occupation distribution
\citep[HOD; e.g.,][]{Zheng:2007lr}.
The mass of dark matter halos can be
determined with HOD models. Many studies have shown that LBGs are strongly clustered
\citep[e.g.,][]{Adelberger:1998qy,Giavalisco:1998uq},
and the brighter galaxies are more strongly clustered at large scales
\citep[e.g.,][]{Adelberger:2005fk,Ouchi:2005vn,Lee:2006qy,Hildebrandt:2007rt}.
In addition, the correlation function of LBGs shows excess power
at small scales ($\theta < 1''$), 
implying multiple galaxies within the same massive dark
matter halo in the context of HOD \citep{Ouchi:2005vn, Lee:2006qy,Lee:2009qy}.
Combined with the UV LF, clustering results also can be used 
to infer the nature of star formation in the LBGs and its dependence on their host halo mass 
\citep{Lee:2009qy}.

In previous deep field surveys, survey areas were 
relatively small.
The largest $z \sim3$ LBG surveys so far with spectroscopic redshifts
are the Keck Baryonic Structure Survey \citep[KBSS;][]{Steidel:2003kx,Steidel:2004fk} and the VLT LBG Redshift Survey 
\citep[VLRS,][]{Bielby:2011qy}.
The KBSS and VLRS 
cover a total area of around 1~deg$^2$ and 3~deg$^2$, respectively, with $\approx2000$ spectroscopic redshifts
\citep{Reddy:2009kx,Bielby:2012uq}.  
The largest coherent structures revealed in these
surveys have sizes comparable to the field size: we clearly have not reached the scale of the largest
structures at that time. The small sample size means that
only simple statistics can be computed, and it is difficult to
sub-divide the sample to probe the dependence of clustering on the 
intrinsic properties of the galaxies. 
Given the difficulty in obtaining even larger spectroscopic samples of faint LBGs, 
the only effective way to expand
the sample size by a large factor is through photometrically
selected samples. For example, the Garching-Bonn deep survey
\citep{Hildebrandt:2007rt} covers $\approx2$ deg$^2$, and $\approx$ 8000
$z\sim3$ photometrically-selected LBGs are selected to study the clustering properties.


The key to establishing a large $z\sim3$ LBG sample is
the availability of deep multi-wavelength imaging, especially deep
$U$ band imaging. The Large Binocular Camera \citep[LBC,][]{Giallongo:2008fk} - 
Blue on the left arm ("DX-side") of the $2\times8.4$~m Large Binocular Telescope (LBT) 
is specially designed to have high throughput
and good image quality in the blue. 
We have carried out a large LBC survey of the NOAO Deep Wide Field Survey \citep[NDWFS,][]{{Jannuzi:1999fr}}
Bo\"otes Field (9 deg$^2$) in the 
$U_{\rm spec}$ band ($\lambda_0 = 3590$\angstrom, FWHM=540\angstrom) and $Y$ band
($\lambda_0 = 9840$\angstrom, FWHM=420\angstrom, Figure~\ref{filters}),
building on the unique
multi-wavelength data set already available for the  Bo\"otes field, while
filling in two critical wavelength gaps. The survey area is about
five times larger than previous studies \citep[e.g. COSMOS, ][]{Scoville:2007fk},  
which allows us
to build a larger LBG sample to further study the luminosity 
function and clustering properties, especially for the brightest
LBGs at redshift $z\sim3$.

This is the first in a series of papers. 
In this paper, we will focus on the photometrically-selected LBGs and
study their UV and NIR LF and clustering properties, especially
for the bright LBGs. In following papers, we will focus on  
spectroscopic confirmation of the most luminous LBGs. 

The paper is organized as follows: observations are discussed in 
section~\ref{observationsection}. 
Data reduction is described in section~\ref{datareductionsection}. 
Sample 
selection is given in section~\ref{selectionsection}. We present our bright end 
rest-frame UV and near-IR LF and stellar mass function (SMF) results in section~\ref{lfsection} and \ref{nir_lf} 
and discuss the evolution of UV LF and SMF with cosmic time in section~\ref{evol}.
Clustering results are presented
in section~\ref{clusteringsection}. Finally, we summarize our results. 
Throughout this paper, we use the following
cosmological parameters for the calculations: Hubble constant, $H_0=70$~km~s$^{-1}$~Mpc$^{-1}$; 
dark matter density, $\Omega_{\rm M}=0.30$; and dark energy density, $\Omega_\Lambda=0.70$
for a flat Universe \citep[e.g.,][]{Spergel:2007fk}.
All the magnitudes are expressed in the AB magnitude system \citep{Oke:1983qy}.



\section{OBSERVATIONS}\label{observationsection}

\begin{figure}[h]
\epsscale{1.2}
\plotone{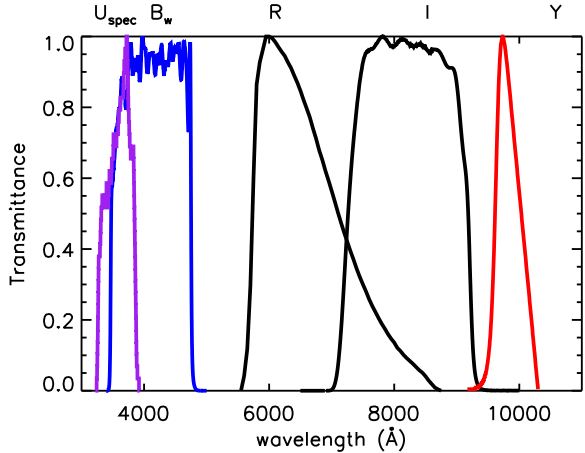}
\caption{Relative transmission curves of the LBC $U_{\rm spec}$-band (purple curve) 
and $Y$-band (red curve). The $U_{\rm spec}$-band and $Y$-band filter curves 
have been corrected by both the CCD Q.E. curve and the atmosphere transmission. 
This plot also shows the transmission curves
of the $B{\rm w}$ (blue curve), $R$ and $I$ bands in the NDWFS Bo\"otes field.
All the transmission curves are
normalized by the peak transmittance for clarity.
\label{filters}}
\end{figure}

\begin{figure*}[ht]
\epsscale{1.2}
\plotone{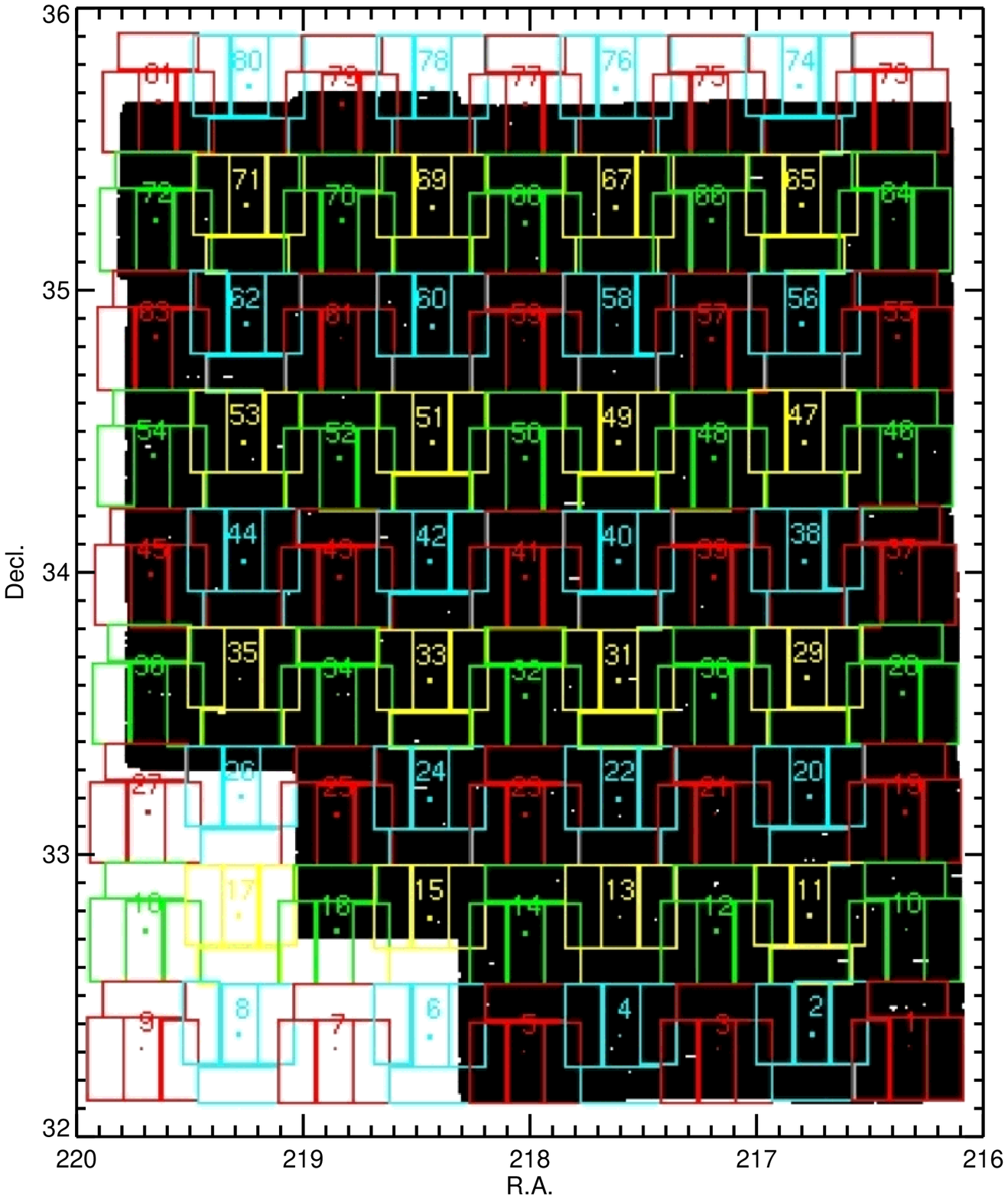}
\caption{$U_{\rm spec}$ coverage map for the Bo\"otes field. The black region is 
covered by the NDWFS $B{\rm w}$, $R$, and $I$ bands. There are a
total of 81 fields designed to cover the whole rectangular region, 63 of
which have overlaps with the NDWFS broad band coverage.
\label{map}}
\end{figure*}

The Bo\"otes Field ($\alpha(J2000)=14^h32^m$, $\delta(J2000)=+34^{\rm o}16'$),
one of the NDWFS fields, 
is a 9 deg$^2$ field covered by a
deep multicolor survey in $B\rm{w}$, $R$, and $I$ optical broad bands \citep{Jannuzi:1999fr},
and $J$, $H$, and $Ks$ near-infrared (near-IR) bands \citep{Elston:2006mz,Gonzalez:2010lr}. 
A shallow $z$-band survey was carried out by \citet{Cool:2007zr}.
Additionally, this deep and wide field has been 
observed at other wavelengths, including in the X-ray with {\it Chandra}
 \citep{Murray:2005ly,Kenter:2005gf,Brand:2006ve},
UV with  {\it GALEX}  \citep{Hoopes:2003qf}, infrared with {\it Spitzer} IRAC  \citep{Ashby:2009kq}
and MIPS  \citep{Soifer:2004pd}, and radio with the 
VLA \citep{Becker:1995lr,de-Vries:2002lr}. 
A redshift survey \citep[the AGN and Galaxy Evolution Survey;][]{Cool:2006bh} has 
also been conducted on this field,
with spectra of roughly 17,000 galaxies and 3,000 AGNs down to I$\approx$20 using the Hectospec instrument
mounted on the 6.5~m MMT.

There are two significant gaps in the optical wavelength coverage in this field:
one is between the $B{\rm w}$ broad band and GALEX NUV, and the other is
between $I$ band and $J$ band. To fill these two gaps,
we have carried out the LBT Bo\"otes field survey with the LBCs mounted on
 the $2\times8.4$~m LBT in binocular mode 
with $U_{\rm spec}$ band ($\lambda_0 = 3590$\angstrom, FWHM=540\angstrom) and $Y$ band
($\lambda_0 = 9840$\angstrom, FWHM=420\angstrom) imaging (Figure~\ref{filters}).
The LBCs are two wide-field cameras, and each is mounted on one of
the LBT prime foci. These two cameras can observe
the same sky field simultaneously. The blue channel 
is optimized for the UV-B bands and the red channel is optimized for the VRIz bands.
The CCD quantum efficiencies are $\approx50\%$ and $\approx10\%$ in the $U_{\rm spec}$
and $Y$ bands, respectively. The CCD pixel size is $0.225''$/pixel.
Each of the cameras consists of four 2k$\times$4k chips, resulting in 
a $23\times23$ arcmin$^{2}$ field of view (FoV). 
Because the layout of the CCD is not a square, the total effective FoV is about 470 arcmin$^2$.

The primary goal of our survey is to use the unique $Y$ and $U_{\rm spec}$ band data to search for
$z\sim7$ quasars at the epoch of cosmic reionization and $z\sim3$ LBGs at the
epoch of the peak in star-forming and quasar activity. In this paper, we focus on the 
$U_{\rm spec}$ band data to establish an LBG sample
and publish the scientific results of the LBG sample.

A total of 81 pointings were designed
to cover the entire rectangular region (Figure~\ref{map}), and 63 of them,
which overlap with the NDWFS optical band coverage by more 
than 50\% of the LBC FoV, were observed. The total survey area is about 8.8 deg$^2$.
For each individual field, a 1200s exposure time observing block is designed. 
The total 1200s exposure time
is divided into five individual 240s exposures with 30$''$ dither patterns. 
The dither pattern allows us to fill up the chip gaps, remove cosmic rays
and bad pixels, and reduce the effects of the bright stars. The
position angles of the neighboring
fields in the declination direction have a 180 deg difference and an $\approx4'\times7'$
overlap region, which allows us to compare the calibration with different position angles.

The observations were carried out in dark time 
from 2008 January to 2009 March in queue observing mode.
There are a total of 718 $U_{\rm spec}$ band science images obtained with a total open 
shutter time of $\approx47.8$ hours. About $30\%$ of the data (222 images) are 
unusable, in which $20\%$ (151 images) are trailed due to motion of the 
telescope during the exposure and $10\%$ (71 images) have
poor image quality (FWHM $>1.8''$). The images with FWHM $>1.8''$
do not make a significant contribution (less than 50\% compared to images
with median FWHM)
to the depth of the final co-added images (see details in 
\S3.4).

The median airmass is 1.11 and the median FWHM is $1.25''$ ($1.33''$)
after (before) removing the bad quality images. The final average effective exposure time
for each individual field is about 30~minutes. 



\section{DATA REDUCTION}\label{datareductionsection}
\begin{figure}
\epsscale{1.2}
\plotone{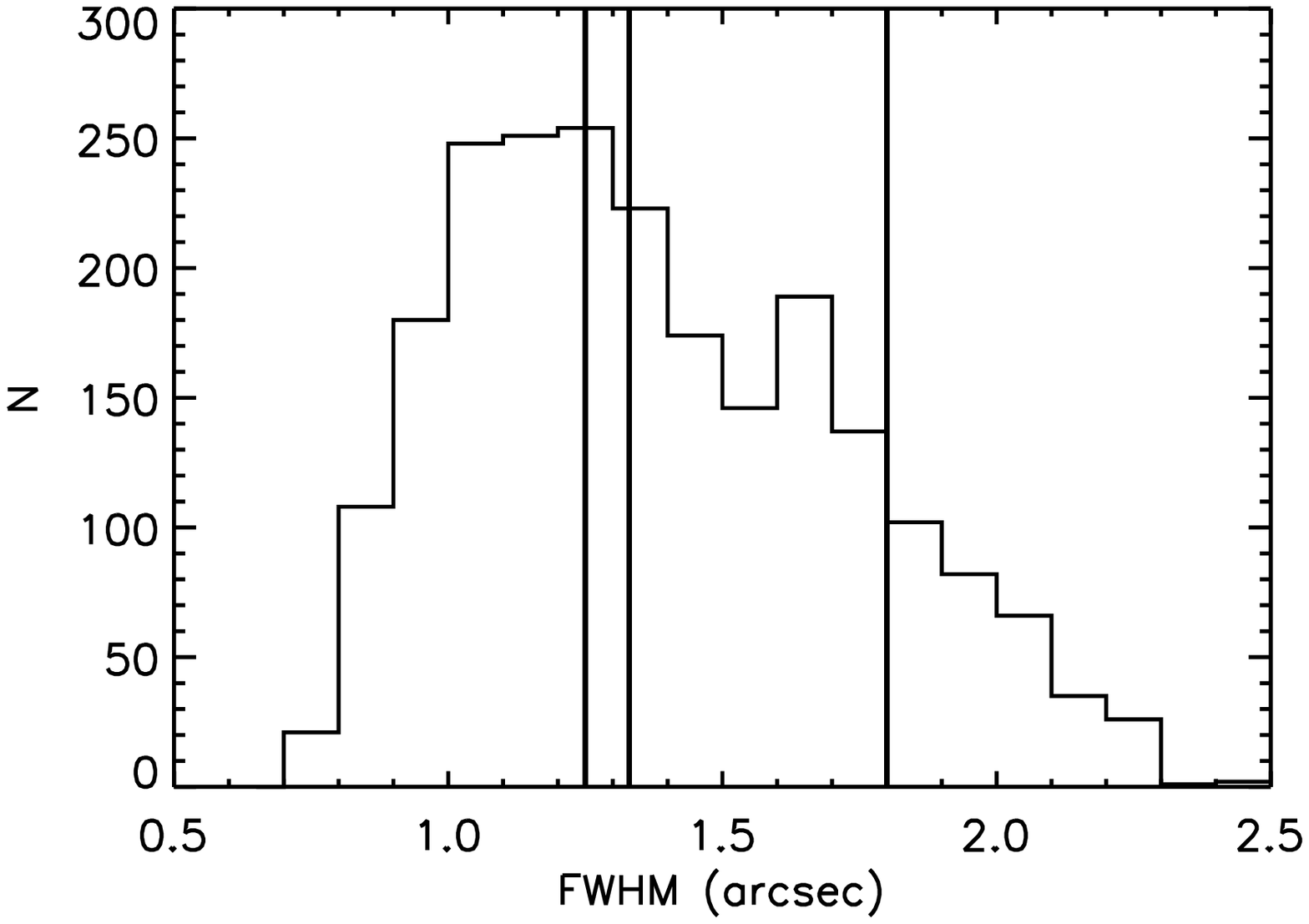}
\caption{Image quality of the $U_{\rm spec}$ band images. The histogram shows
the distribution of the FWHMs of each chip of the LBC. The median of the FWHM is $1.33''$/$1.25''$ 
before/after removing the bad image quality images ($\rm FWHM>1.8''$). The three vertical lines mark
the FWHMs of $1.25''$, $1.33''$, and $1.8''$ from left to right. 
\label{fwhm}}
\end{figure}

\begin{figure}
\epsscale{1.2}
\plotone{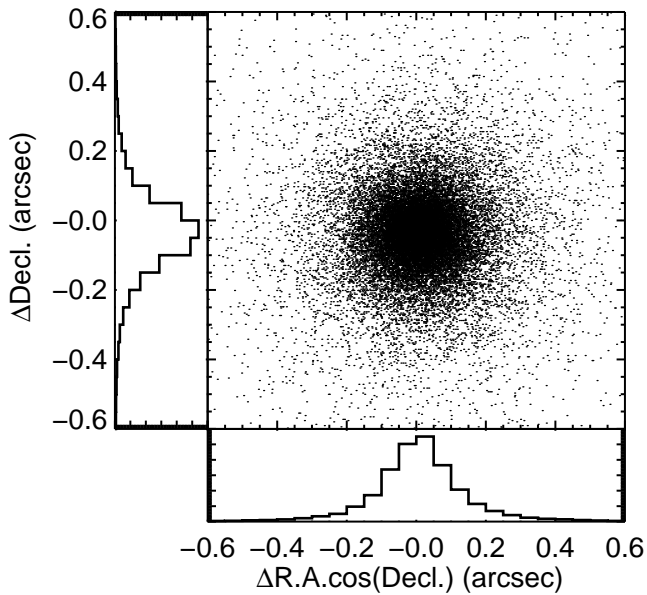}
\caption{Accuracy of the astrometry. The distribution of the difference of R.A.$\times\cos$(Decl.)
and Decl. between LBC $U_{\rm spec}$-band images and SDSS images. The FWHMs of both distributions
are about $0.2''$, which corresponds to a $1\sigma$ astrometric uncertainty of $0.08''$.
\label{astrometry_23}}
\end{figure}

\begin{figure}
\epsscale{1.2}
\plotone{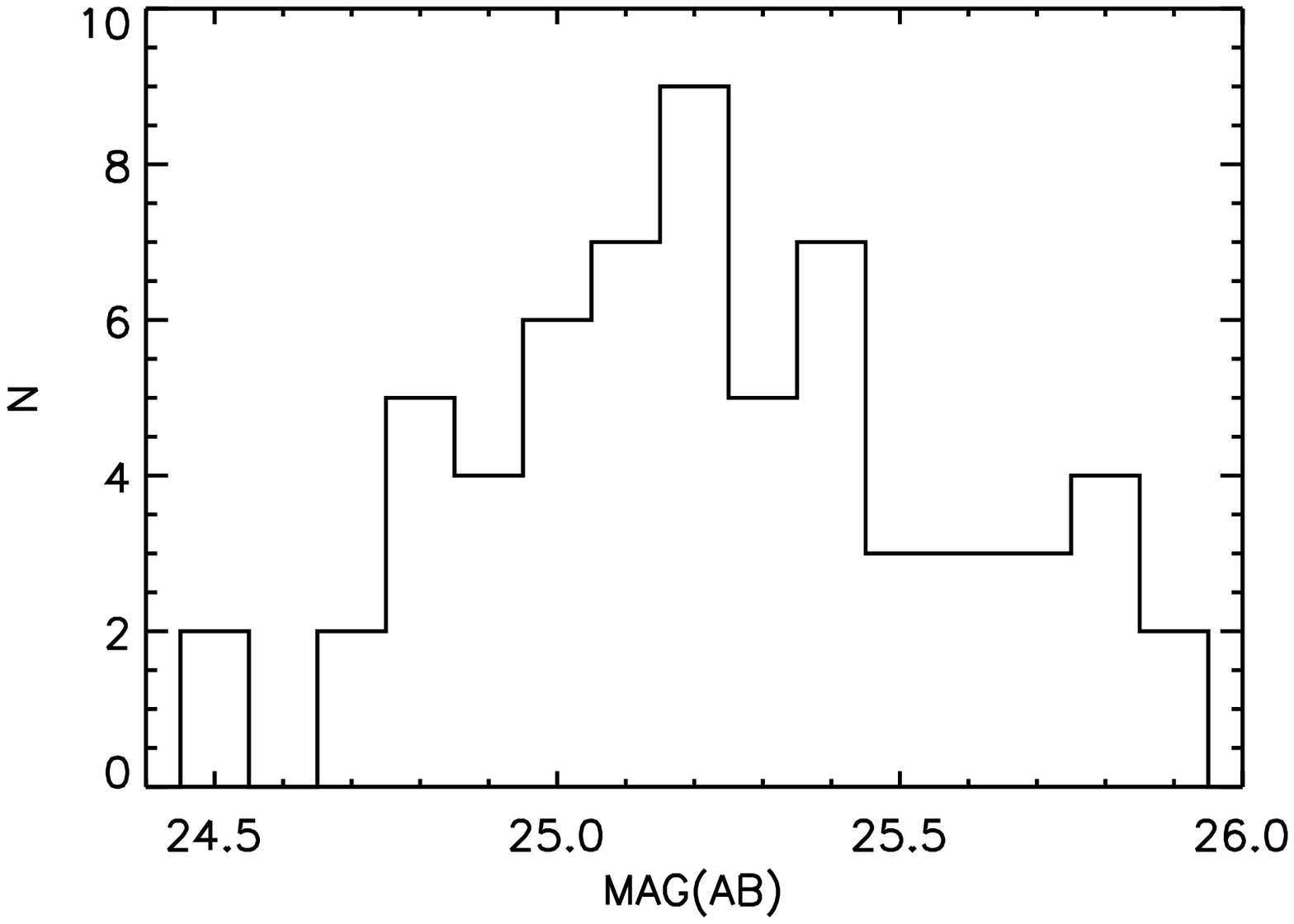}
\caption{Distribution of magnitude with $5\sigma$ detections for the $U_{\rm spec}$ band. The median
depth is 25.2 mag. \label{maglimit}}
\end{figure}

\subsection{Bias and Flat Correction}
All images are inspected visually to remove 
elongated images due to tracking and guiding issues during observing
before further processing. About 20\% of images are removed 
in this step.


Data are reduced using custom IDL routines. 
The bias level measured from the overscan region is subtracted
from individual bias frame images, which are then combined to
construct the master bias frame. Super-sky flats
are created by combining the science frames in each
observation run, scaling by the mean value with $3 \sigma$ clipping
to remove cosmic rays and objects. Science images are overscan
and bias subtracted, and then divided by the super-sky flats to correct
the CCD response. Finally, cosmic rays in the images were identified
rejected using an identification algorithm based on 
Laplacian edge detection \citep{van-Dokkum:2001lr}.

\subsection{Image Quality Measurement}

We estimate the FWHM of each scientific frame
as follows: (1) a few stars across 
the field are selected, and the FWHM of the stars is measured using {\it imexamine} 
in IRAF\footnote{
IRAF is distributed by the National Optical Astronomy Observatories,
which are operated by the Association of Universities for Research
in Astronomy, Inc., under cooperative agreement with the National
Science Foundation.} as an input for the next step; (2)
a catalog for each field is created by SExtractor \citep{Bertin:1996fk} 
with the parameter of {\em SEEING$_-$FWHM} set to the value
from step 1. All objects with six or more connected pixels with 
flux 3.0 times greater than the sky noise are detected. Well-detected bright
stars ($18.5<{mag_{auto}}<20.0$ and $star_-class>0.95$) are selected from the SExtractor
output catalog; and (3) image coordinates of well-detected bright stars are used as the 
input information for the FWHM measure task, {\em psfmeasure}, in IRAF. This task fits the bright
stars profile as a Moffat distribution function. The median of FWHMs of the best-fit Moffat profile
in each frame is used to represent its image quality.

Figure~\ref{fwhm} shows the $U_{\rm spec}$ band image quality. 
Frames with FWHM larger than 1.80$''$ were not used for further analysis and co-addition.
After removing the bad image quality images (FWHM$>1.8''$), the median FWHM  is $1.25''$,
and the first and third quartiles are $1.07''$ and $1.50''$ respectively.
%

\subsection{Astrometric Calibration}\label{astrometry}

The catalogs created by SExtractor in the previous section
are used as an input catalog for SCAMP (Software for Calibrating AstroMetry and Photometry) 
 \citep{Bertin:2006qy}. 
To compute the astrometric solution, 
we use well-detected objects that meet the following
criteria: (1) the object is not saturated; (2) the $S/N$ of the object is greater than 10; and
(3) the FWHM of the object in the SExtractor output catalog is between $2''$ and $10''$.
These well-detected objects from the 
input catalog are used to search for matches in the SDSS-DR6
catalog within a 3$''$ radius. We first use a linear model 
with only an $x$, $y$ directional offset and without rotations to 
obtain a rough astrometry solution. 
Based on this solution, we then 
use a third order polynomial to fit the $x$, $y$ offset and the
rotation to get a refined solution.
Using this procedure, an accurate astrometric 
solution is derived for each field with the $1\sigma$ uncertainty less
than $0.1''$ in both R.A. and Decl. direction (Figure~\ref{astrometry_23}). 

Previous studies \citep[e.g.,][]{Cool:2007zr} have shown that there
is $\approx0.3''$ offset between the NDWFS catalog\footnote{\tt http://www.noao.edu/noao/noaodeep/DR3/dr3cats.html}
astrometry {
and the SDSS astrometry. 
Thus, we also register the NDWFS $B{\rm{w}}$, $R$, and $I$ band images
to the SDSS-DR6 catalogs using the same
method.

\subsection{Image Co-addition}

Before co-adding images, we subtract the sky
background from the science frames and generate
a weight map for each frame.
The background is constructed from the {\em -object} image 
created by SExtractor, in which 
the detected objects have been subtracted from the image.
First, the {\em{-object}} image is divided into $\approx 100$ background mesh
regions with the size of 130~pixel$\times$130~pixel.
Then the median background is computed for each region,
and it is fitted with a second order polynomial and subtracted
from the science images.
The weight map value is computed as follows:
\begin{equation}
w = \frac{1}{\rm FWHM^2\sigma^2},\label{weight}
\end{equation}
where the FWHM is described in \S 3.2, and $\sigma^2$ is 
the sky variance.

SWarp \citep{Bertin:2002uq} is used to co-add images for each field.
First, the input science images and weight maps are
re-sampled to a common pixel grid. The interpolation function we use to re-sample 
images is 
LANCZOS3, a $\prod_d$ sinc($\pi x_d$)sinc($\frac{\pi}{4} x_d$)
response function with ($-3 < x_d \le 3$).
The output co-added image 
 is a weighted average of input values after 3 $\sigma$ clipping:
\begin{equation}
F = \frac{\sum_i w_i f_i}{\sum_i w_i},
\end{equation}
where $w_i$ is the weight of the pixel from re-sampled
weighted map, and $f_i$ is flux value of the pixel from the 
science image. 
The output co-added weighted map is the sum of input weights:
\begin{equation}
W = \sum_i w_i.
\end{equation}
Finally, we create exposure maps to record 
the exposure time for each 
pixel in the co-added images.


\subsection{Photometric Calibration}
The imaging data are calibrated with SDSS data release 6 (DR6) photometry. 
The SDSS $u$ band transmission curve is similar to LBC $U_{\rm spec}$,
with $\lambda_0 = 3540${\angstrom} and FWHM = 570{\angstrom}. 
Bright stars ($18.0 < u<19.5$) in the NO. 57 field which was taken
in photometric conditions are used to determine an offset between
the LBC $U_{\rm spec}$-band and SDSS $u$-band images, and the color term coefficient.
We find the color term is very small, about $-0.01\times(u-g)$.
 Bright stars (star$_-$class $>$ 0.9 and magerr$_-$aper($8^{\prime\prime}$) $<$ 0.02)
in the overlap regions are used to determine the offset of  the photometric zeropoint in other fields.
These stars are also used to check the magnitude difference between two 
neighboring fields. As the position angle of the neighboring frame is offset by 
180 deg, this check will give us the upper limit of the magnitude uncertainty from
the calibration.
The average standard deviation of the difference is $\approx0.05$,
implying the photometric accuracy is $\approx0.05$. 
Another way to check the photometric accuracy is to compare magnitudes of
stars within the same field observed in 
different individual exposures; from this check we find the standard deviation 
is about 0.04,
therefore, the $U_{\rm spec}$-band magnitude uncertainty in the Bo\"otes survey is $\approx0.04-0.05$.
The magnitude uncertainty is mainly introduced by the flat field and the non-uniformity of the image quality across the field.

{

\subsection{Survey Depth}
We use the following process to determine the $5\sigma$ detection depth in the 
$U_{\rm spec}$-band image for each field:
(1) we generate five catalogs for each individual field, and
each catalog has 6,000 simulated stars with magnitudes between 22.0 and 28.0.
The Moffat profile with the FWHM value the same as the image quality of each individual field is used to generate the 
light profile of stars, and these fake stars are added to the real $U_{\rm spec}-$band images; and
(2) we use the SExtractor to detect and measure
the simulated stars in the same manner as for the primary catalogs. 
The standard deviations between the measured
magnitude and the input magnitude are calculated in different magnitude bins. 
For each field, the magnitude bin in which the standard deviation is 0.2 is 
considered as the magnitude limit with $5\sigma$ detection.  
The median depth is around 25.2 AB magnitude, and the first and third
quartiles are 25.0 and 25.4 AB magnitude, respectively.
The distribution of the $5\sigma$ 
limit magnitude is shown in Figure~\ref{maglimit}.

In this paper, we also use the $B_{\rm W}$, $R$ optical 
broad band images\footnote{ftp://archive.noao.edu/ndwfs/dr3/} 
taken by the NOAO/KPNO Mosaic-1 (8K$\times$8K CCD) Wide Field Imager 
on 4-m Mayall Telescope at Kitt Peak National Observatory (KPNO) covering
the entire field. The typical exposure times in these two bands are $\approx8400$s and
$\approx6000$s, respectively. The median image qualities of both $B_{\rm W}$- and $R$-band images are 1.11.        
The median 5-$\sigma$ depths are
26.3 and 25.3 AB magnitude  in $B_{\rm W}$ and $R$ bands, respectively (see Table~\ref{depth}).
The median image qualities 
and 5$\sigma$ depths of $B_{\rm W}$- and $R$-band images
are adopted from the third data release of the NDWFS\footnote{http://www.noao.edu/noao/noaodeep/DR3/dr3-data.html}.

\section{LBG SAMPLE SELECTION}\label{selectionsection}
\begin{figure*}
\epsscale{1.1}
\plottwo{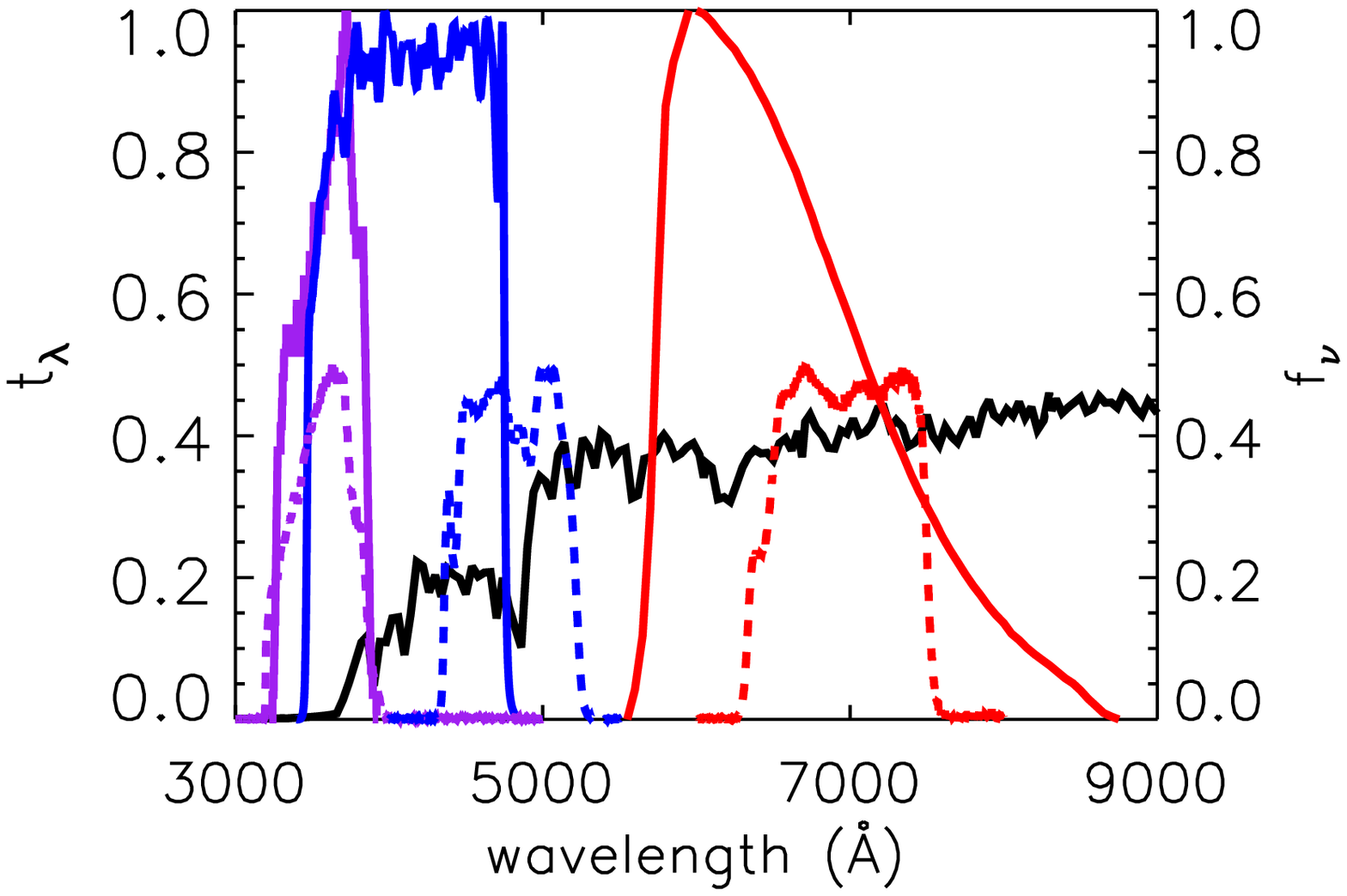}{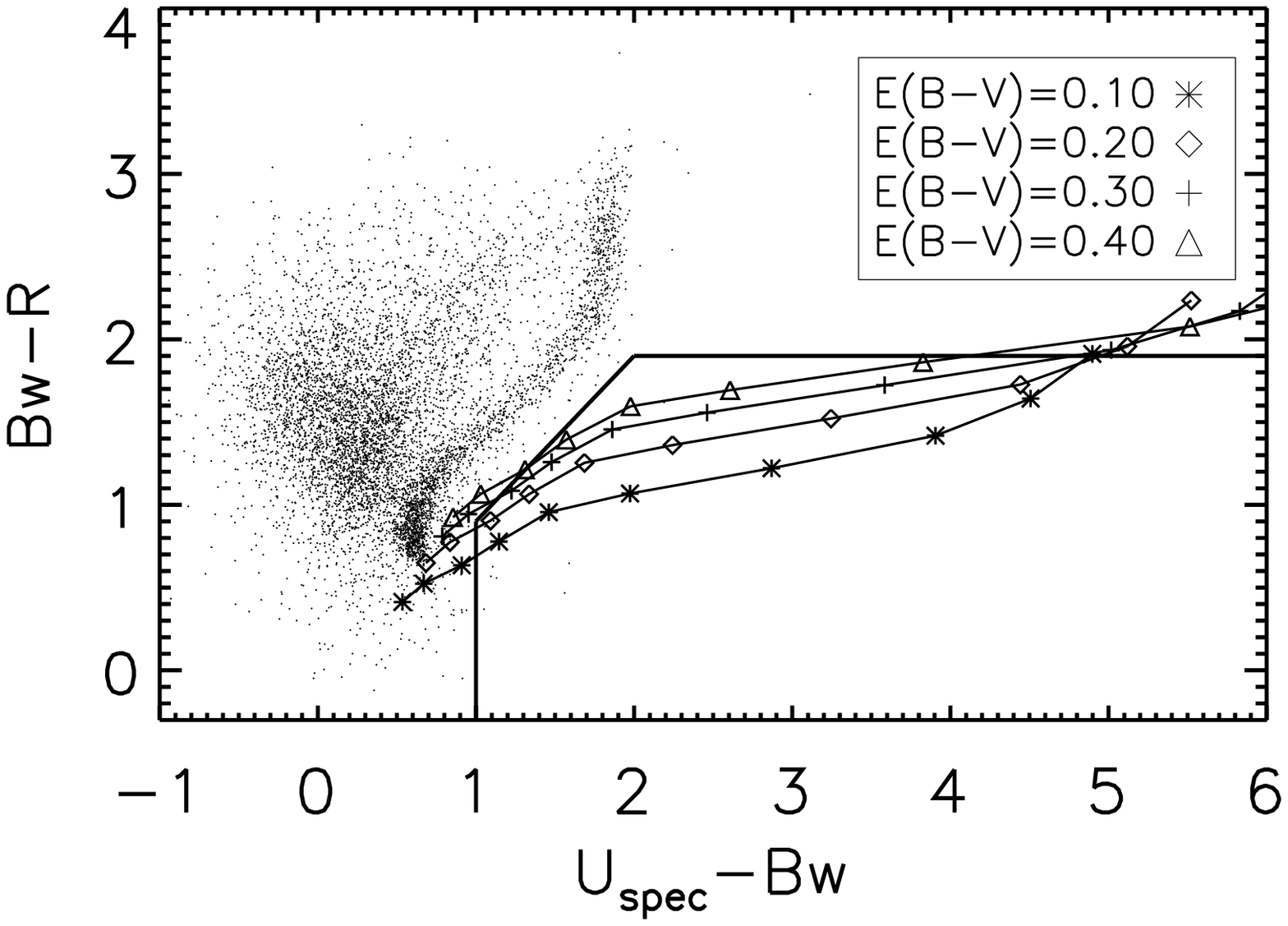}
\caption{Left panel shows the $U_{\rm spec}$ (purple solid curve), $B_{\rm
W}$ (blue solid curve), and $R$ (red solid curve) broad band filters relative transmission curves
and the spectrum of a model of a 300~Myr old star-forming galaxy with a constant SFR at
$z\sim3$ (black curve). For comparison, we also plot the $U_{\rm n}$ (purple dashed curve), $G$ (blue dashed curve), and 
$R$ (red dashed curve) broad band filters used in \citet{Steidel:2003kx}. For clarity, we scale
the peak of the transmittance as 1.0 for the filters used in this work and as 0.5 for the filters
used in \citet{Steidel:2003kx}.
The right panel shows the color evolution of the model galaxy as a function of 
reddening and redshifts in the $B{\rm w}-R$ vs. $U_{\rm spec}-B{\rm w}$
diagram. The left-most points are for redshift
$z\sim2.5$. The point step corresponds to intervals of $\delta
z=0.1$. The region enclosed by the solid line indicates the selection
criteria. The dots are the stars that are well detected ($\rm mag_-err<0.05$) in 
all three bands, which show a tight stellar locus.
\label{simulation}}
\end{figure*}

\subsection{Photometry}
To select LBGs, the crucial step is to 
find the $U$-dropout objects. In some cases, the $U$-dropout objects cannot
be directly detected and measured on $U$-band images, therefore,
the NDWFS $R$-band images are used as the detection images
in this study. 

First, we map the NDWFS $R$-band images using SWarp
to the LBT $U_{\rm spec}$-band images with the same pixel size and image size. 
Then we use SExtractor in double-image
mode with the mapped $R$-band images as detection images and $U_{\rm spec}$-band images as
measurement images. A source is 
considered to be detected 
if the number of connected pixels with flux 0.6 times greater than the 
sky $\sigma$ exceeds four pixels after the original image is convolved with a 
$9\times9$ convolution mask of a Gaussian point-spread function with FWHM = 5.0 pixels in the $R$-band image.
We reduce the NDWFS $B_{\rm W}$ data in the same manner. 
The astrometry has been registered to the SDSS-DR6 catalog as discussed in \S \ref{astrometry}. 
 
The $U_{\rm spec}$-band exposure map masks are used as the external flags to 
obtain the exposure time for each object. Objects
with $U_{\rm spec}$-band exposure times less than 720s are ignored.
The aperture magnitude (mag$_{-}$aper) with aperture size of $2.0\times$FWHM is then used for 
color selection.
The aperture 
correction is estimated for each field by applying an 8$''$ aperture to 
measure the total flux of bright stars. Then the aperture corrections (i.e. 
$\Delta$mag = mag$_{-}$aper(8$''$)-mag$_{-}$aper(2$\times$FWHM)) --
with values around $-0.2$ -- are used
to correct the flux loss due to measurement in relatively small apertures.
For those $U_{\rm spec}$ non-detected sources, we set the magnitude upper limits at 
$1\sigma$.
Sources from the $U_{\rm spec}$, $B_{\rm W}$ and $R$-band catalogs within 1$''$ positional variation
are matched together to generate the $U_{\rm spec}-B_{\rm W}$ versus $B_{\rm W}-R$ color-color
diagram.

\subsection{Sample Selection Criteria}\label{criterion}
To determine the LBG sample selection, the BC03 standard stellar synthesis population 
model \citep{Bruzual:2003lr} is used to build a series of  
spectral templates of star-forming galaxies. We adopt a spectral model
with a constant SFR, 
a Salpeter initial mass function \citep[IMF;][]{Salpeter:1955kx}, one solar metallicity 
abundance, and an age of 300~Myrs to simulate the spectra
of star-forming galaxies. This model will give us a typical intrinsic
LBG spectral energy distribution \citep[SED; e.g.,][]{Steidel:2003kx}.

The templates are modified by the
intergalactic medium (IGM) absorption model of \citet{Madau:1995lr} and 
reddened using the attenuation law of \citet{Calzetti:2000vn} with
reddening of $E(B-V)=0.0-0.4$. Then the wavelength of 
the spectra is shifted by a factor of $1+z$ to derive the spectra in
the redshift range from 2.5 to 3.6. 
Figure~\ref{simulation} shows one of the star-forming galaxy 
template spectra with $E(B-V)=0.2$ at a redshift of $z=3$, and how the galaxies with
given $E(B-V)$ evolve in the $U_{\rm spec}-B_{\rm{W}}$ vs. $B_{\rm W}-R$ color-color diagram 
with redshift. The left-most  point for each track represents a redshift
of $z=2.5$ and each step corresponds to a redshift interval of 0.1. The lower redshift
limit is primarily determined by the  $U_{\rm spec}-B_{\rm W}$ cut, and the upper redshift limit
is mainly determined by the  $B_{\rm W}-R$ cut. We also plot the well detected stars ($\rm mag_-err<0.05$) in 
all three bands in the color-color diagram. It shows a stellar
locus, which is well separated from the tracks of the $z\sim3$ LBGs. To select the
LBGs with $2.7 < z <  3.3$ and maintain enough separation from the stellar locus 
to reduce the contamination rate from stars, the following selection
criteria for $z\sim3$ LBGs are adopted:
\begin{eqnarray}
 U_{\rm spec}-B{\rm w} > 1.0, \nonumber \\
 B{\rm w}-R<1.9, \nonumber \\
 B{\rm w}-R < U_{\rm spec}-B{\rm w}+0.1, \nonumber \\
 R<25.0 .\label{criterionequation}
\end{eqnarray}


\subsection{Selection Function}\label{completenesssection}
\begin{figure}
\epsscale{1.2}
\plotone{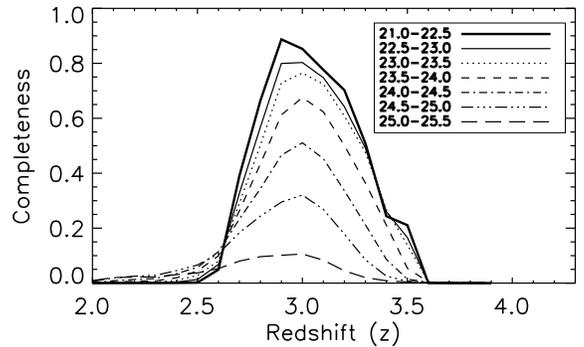}
\caption{Selection completeness as a function of redshift in different $R$-band magnitude bins.
\label{completeness}
}
\end{figure}

\begin{figure}
\epsscale{1.2}
\plotone{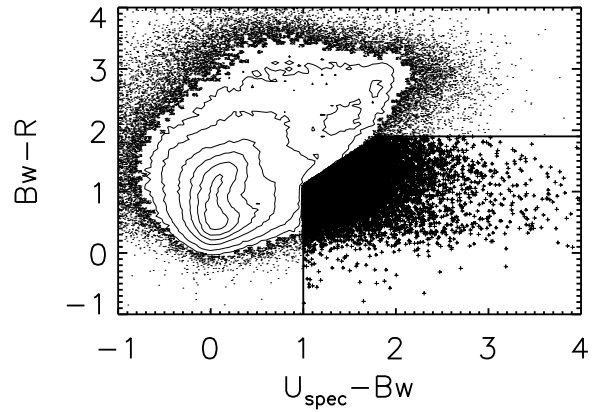}
\caption{$B_{\rm W}-R$
versus $U_{\rm spec}-B_{\rm W}$ color-color diagram. The photometrically-selected LBGs at $z\sim3$ are
selected in the region bounded by the solid line. A total of
$14,485$ photometrically-selected LBGs are selected as U-dropouts. The
crosses represent the selected LBGs. The reason for the
sharp edge at $B_{\rm W}-R = 1.9$ and $U_{\rm spec}-B_{\rm W} = 1.0$ is that 
we do not plot the $U_{\rm spec}$ band undetected
sources without falling in the selection criterion region.\label{color}}
\end{figure}

For a well detected galaxy with intrinsic $U_{\rm spec}-B_{\rm W}$ and
$B_{\rm}-R$ colors right in the color-color selection region,
the probability to select this galaxy as an LBG is mainly influenced by
the range of intrinsic SEDs of the population at that redshift, some of which could
scatter the observed colors beyond
the selection region boundaries or band detection limits.
In this paper, the completeness is derived from the selection function,
which describes the detection probability ($P(m, z, {\rm SED}$)) of an LBG spectral template
with a given redshift, magnitude and 
SED falling within the selection criteria.

The procedures to calculate the LBG selection function are as follows:
the spectral templates generated in section~\ref{criterion} are used to derive the intrinsic
color distribution. For galaxies with a constant SFR, Salpeter IMF, 
solar metallicity, and 300~Myr age, as is typical for an LBG SED \citep[e.g.,][]{Steidel:2003kx},
the SEDs of the galaxies are influenced only by the reddening $E(B-V)$.
Therefore, in this case, $P(m, z, {\rm SED})$ is equivalent to 
$P(m,z, E(B-V))$. The detection probability is then a function of magnitude, redshift and dust extinction. 
The $E(B-V)$ distribution of our sample is taken from the results of \citet{Reddy:2008fj} Table 5, 
which is from -0.1 to 0.4. 
A series of SEDs are generated with $2.0 < z < 4.0$ ($\Delta z=0.1$) and $E(B-V)$
from -0.1 to 0.4 ($\Delta E(B-V) = 0.1$). 
The broad band colors are fixed for a given magnitude, dust extinction and redshift combination.
The expected colors of $U_{\rm spec}-B_{\rm W}$ and 
$B_{\rm W}-R$ for a given redshift and $E(B-V)$ are derived by convolving the spectral template with the filter curves.
Ten thousand simulated objects, following expected $U_{\rm spec}-B_{\rm W}$ and
$B_{\rm W}-R$ colors, are put on the $U_{\rm spec}$, $B_{W}$, and $R$ images for each $R$ magnitude
($\Delta$mag$=0.5$), redshift ($\Delta z=0.1$), and $E(B-V)$ ($\Delta E(B-V) = 0.1$) bin, based on their expected noise characteristics on the real images.  The size of faint LBGs is compact with $r_e<0.5''$ \citep[e.g.,][]{Ferguson:2004lr}, which cannot be resolved in our
ground-based images, therefore, we use the Moffat profile to simulate the light distribution of the LBGs.
The method of detection and measurement of these simulated objects is the same as that used for our real objects.
Then the $P(m,z, E(B-V))$ is derived from the fraction of the simulated objects meeting 
the selection criteria. By weighting $P(m, z, E(B-V))$ with 
the distribution of the $E(B-V)$ \citep{Reddy:2008fj}, the selection function,
i.e., the LBG detection probability, as a function of redshift for a given 
magnitude bin is finally determined.  

Figure~\ref{completeness} shows the selection function as a function of redshift in
different $R$-band magnitude bins from 21.0 to 25.0. 
The redshift range is $2.7<z<3.3$ with a peak at $z=2.9$.
The completeness decreases for fainter galaxies, because the magnitude error scatters color-color points
out of the selection region and the detection completeness for $R$ band
drops very quickly for the faint end (e.g., $R=24.5-25.0$).  For the
fainter magnitude bins, low redshift galaxies have a greater chance to be
scattered into the selection 
region due to the large magnitude errors. Thus the redshift distribution of faint galaxies shows an extended tail at the low redshift end.

\subsection{A Sample of  Photometrically-Selected $z\sim 3$ LBGs}
\begin{figure}
\epsscale{1.25}
\plotone{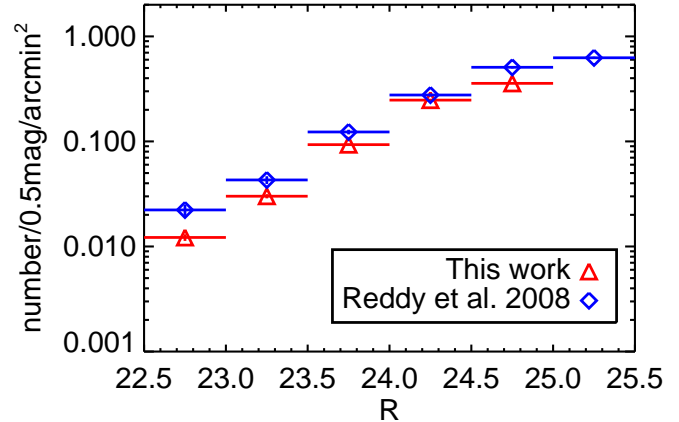}
\caption{Surface density of $z\sim3$ LBGs as a function of $R$-band magnitude
in this work (red triangles) and in \citet[blue diamonds]{Reddy:2008fj}.}
\label{sd}
\end{figure}

Figure \ref{color} shows the distribution of objects from the Bo\"otes field 
in the $U_{\rm spec}- B_{\rm W}$ vs. $B_{\rm W}-R$ color-color diagram. 
Since we do not require the object to be well detected ($10\sigma$ detection)
in all three bands as we did in Figure~\ref{simulation}, the stellar locus and galaxy 
distribution have greater scatter than those in Figure~\ref{simulation}, which contributes to the contamination of the LBG sample.
Using the selection criteria discussed in \S 4.2, a total of $14,485$ photometrically-selected LBGs (cross
symbols in Figure~\ref{color}) are selected down to $R=25.0$ in
the 8.8 deg$^2$ area, leading to an LBG surface density $\Sigma= 0.47\pm0.03 $ galaxies arcmin$^{-2}$.
This value is smaller than the result,  $\Sigma \sim1.8$ galaxies arcmin$^{-2}$, 
in \citet{Steidel:2003kx}, which is 0.5 magnitude deeper than our survey.
Figure~\ref{sd} shows that the surface number density of the $z\sim3$ LBGs in this work
is systematically lower than that in \citet{Reddy:2008fj}. The low number density is mainly due to the 
narrower redshift selection function in our sample compared to that in \citet{Reddy:2008fj}. Our shallower
survey also increases the photometric errors and decreases the detection rate for a given magnitude, which reduces the completeness.

\section{UV LUMINOSITY FUNCTION}\label{lfsection}

\begin{figure*}
\epsscale{1}
\plotone{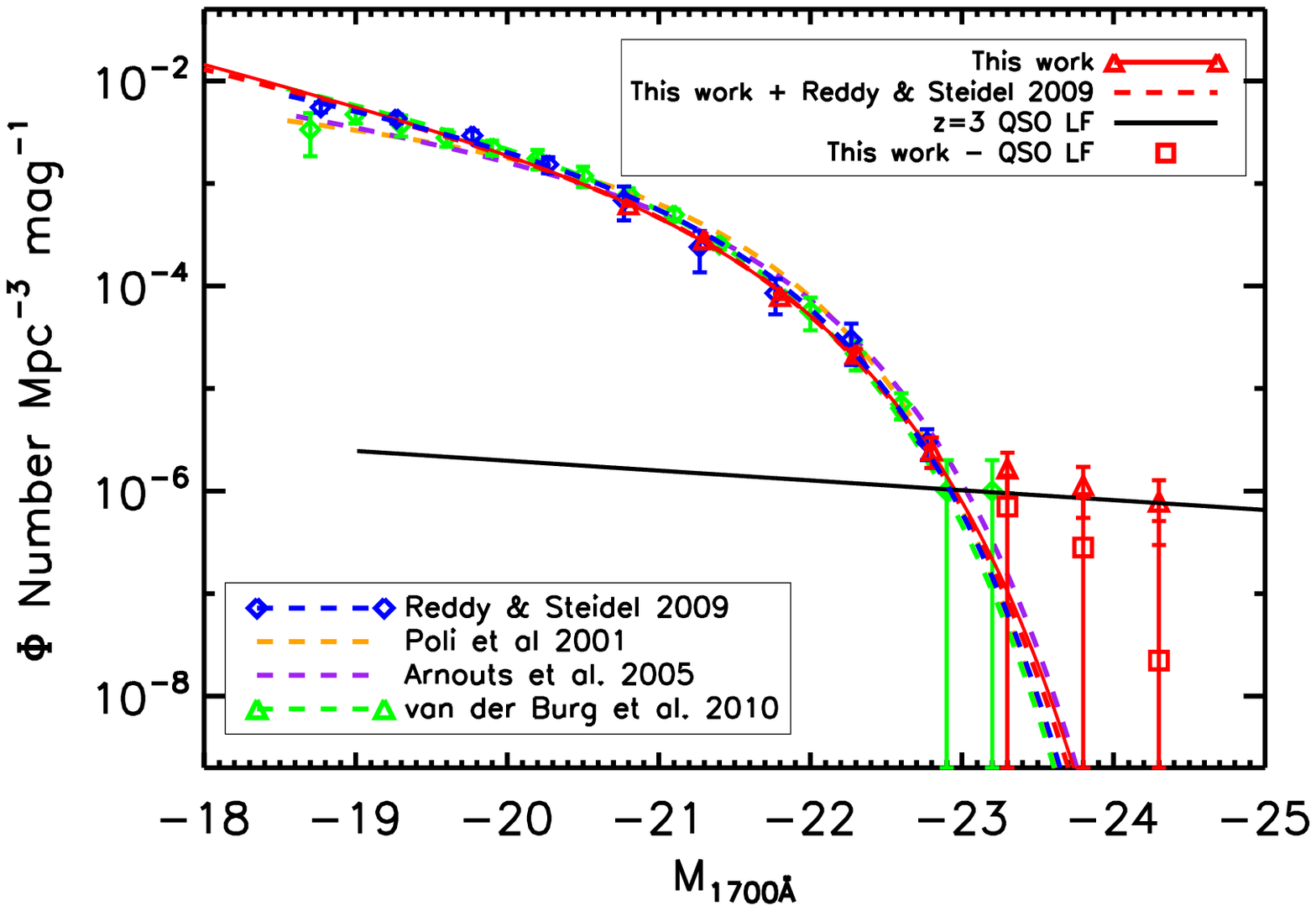}
\caption{Rest-frame UV (1700\rm\angstrom) luminosity function (LF) of the $z\sim3$ LBGs in Bo\"otes field (red triangles)
and the best-fit Schechter function (red dashed and solid curves). For comparison, we also plot
the LFs of $z\sim3$ LBGs from \citet[][the blue diamonds and the dashed curve]{Reddy:2009kx}, 
\citet[][orange dashed curve]{Poli:2001kx}, \citet[][purple dashed curve]{Arnouts:2005yq}, and 
\citet[][green triangles and dashed curve]{van-der-Burg:2010vn}.
All these works are consistent with each other at the luminosity range, $-22.5<M_{1700\angstrom}<-20.5$.
The bright end of the LF in this work shows excess compared 
with the Schechter function. The excess can be explained by the $z\sim3$ quasar LF \citep[black solid line;][]{Hunt:2004fk}.
The red squares represent the residual values that quasar LF is subtracted from the LBG LF measurements.
\label{lf}
}
\end{figure*}

In this section, we will measure the rest-frame UV LF of LBGs
based on their $R$-band magnitudes.


\subsection{Results}\label{lfresults}
The $R$-band filter ($\lambda_{\rm{eff}}=6407\rm{\angstrom}$)  traces the  
rest-frame UV ($\sim1700$\angstrom) for this LBG sample with a mean redshift of 
$z\sim 2.9$. The following formula is used to convert the apparent 
$R$-band AB magnitude ($m_R$) to the absolute magnitude at rest-frame 1700$\rm{\angstrom}$ 
($M_{1700\rm{\angstrom}}$),
\begin{eqnarray}
M_{1700{\rm{\angstrom}}} &=& m_R-5\log_{10}(d_{\rm L}/10{\rm pc})+2.5{\log}_{10}(1+{z}) \nonumber \\
&+&\left(m_{1700}-m_{\lambda_{\rm obs}}/(1+z)\right),
\end{eqnarray}
where $d_{\rm {L}}$ is the luminosity distance in pc and $z$ is the redshift.
The second and third terms of the right side are the distance modulus, and 
the fourth term is the K-correction between rest-frame 1700{\angstrom} and the 
$R$-band, which is about 0 \citep{Sawicki:2006fj}.
From the simulation in section~\ref{completenesssection}, the
mean redshift of the LBG sample is 2.9, which corresponds
to a distance modulus of 45.46.

The  rest-frame 1700$\rm{\angstrom}$ LF ($\Phi(M_{1700\rm{\angstrom}})$) 
and its statistical uncertainty in a given magnitude bin can be computed based on \citet{Schmidt:1968fj}:
\begin{equation}
\Phi(M_{1700{\rm\angstrom}}) = \frac{1}{\Delta m}\frac{N(1-f)}{V_{\rm{eff}}},
\end{equation}
and 
\begin{equation}
\Delta\Phi(M_{1700{\rm\angstrom}})=\frac{1}{\Delta m}\frac{\sqrt{N(1-f)}}{V_{\rm{eff}}},
\end{equation}
where $\Delta m$ is the magnitude bin size, which is 0.5 in this paper, $N$ is 
the number of U-dropout LBG candidates falling into the magnitude bin, and
$f$ is the fraction of contamination for the magnitude bin.
In this paper we do not have any spectral observations of these candidates.
Therefore, we adopt the contamination fraction, $f$, from Table~3 in \citet{Reddy:2008fj}.
The value of $f$ is about 0.7 at the bright end and less than 0.01 at the faint end.
We will discuss how the contamination rate affects our LF measurements later in \S5.3.
$V_{\rm{eff}}$ is the effective comoving volume in units of Mpc$^3$.

In a flat Universe, the comoving volume per solid angle per redshift can be calculated as
\begin{equation}
\frac{dV}{d\Omega dz} = \frac{c r(z)^2}{H(z)},
\end{equation}
where $r(z) = \int\frac{c dz}{H(z)}$. 
The effective comoving volume (V$_{\rm eff}$) can be calculated from the comoving volume,
\begin{equation}
V_{\rm {eff}} = \int_{\Delta z}\int_{\Delta m} dz dm P(m,z) \frac{V(z)}{dz dm}\label{veff},
\end{equation}
where $P(m,z)$ is the completeness of the sample as a function of redshift ($z$) and 
$R$-band magnitude ($m$), which has been determined in \S\ref{completenesssection}.
The effective volume for a given magnitude is computed by integrating
equation~\ref{veff} using the results of the selection function $P(m,z)$. 
Then the LF is calculated for each magnitude bin. 

The red triangles in Figure~\ref{lf} represent the UV LF measurement result of our LBG sample.  
The UV LF is fitted by the Schechter function:
\begin{eqnarray}
&&\Phi(M_{1700{\rm \angstrom}})\rm dM(1700{\rm \angstrom}) = \nonumber \\
	&& \frac{2}{5}\Phi^{\star}\ln(10)[10^{\frac{2}{5}(M^{\star}-M)} ]^{\alpha+1}\exp[-10^{\frac{2}{5}(M^{\star}-M)}]\rm dM, 
	\label{schechter}
\end{eqnarray}
and the best-fit parameters for the Schechter function are $\Phi^{\star}=(1.06\pm0.33)\times10^{-3}$ Mpc$^{-3}$, $M^{\star}=-21.11\pm0.08$, 
and $\alpha=-1.94\pm0.10$. For the LF fitting, we do not use the LF in the magnitude bins
brighter than $M_{1700\rm\angstrom}=-23$, as they are significantly overestimated due to contamination by quasars.
The survey depth of the Bo\"otes field is only about 0.5 magnitude fainter than  $M^{\star}$, thus these data 
cannot be used to constrain the faint end slope of the LF very well. Therefore, we combined our
LF measurement with the LF at $M({1700{\rm \angstrom}})>-20.5$ from \citet{Reddy:2009kx}
to fit the Schechter function (\ref{schechter}). We find that the best-fit parameters are 
$\Phi^{\star}=(1.12\pm0.17)\times10^{-3}$ Mpc$^{-3}$, $M^{\star}=-21.08\pm0.05$,  and $\alpha=-1.90\pm0.05$.  
Combining the LF measurements from different data sets could bring significant systematic errors into the LF fitting due to 
the quite different filter sets and selection criteria.
In Figure~\ref{lf}, we compared the LF derived 
 in this work with those in 
\citet{Reddy:2009kx}, \citet{Poli:2001kx}, \citet{Arnouts:2005yq}, and \citet{van-der-Burg:2010vn}
in the magnitude range $-23<M({1700{\rm \angstrom}})<-20.5$.
They are consistent with each other
within 1$\sigma$. 

\subsection{UV Luminosity Density}
The UV luminosity density from integrating the Schechter function for
a faint luminosity limit is given by:
\begin{equation}
\rho_{L_{\rm UV}} = [\Gamma(\alpha+2)-\gamma(\alpha+2,L_{\rm lim}/L^{\star})]\Phi^{\star}L^{\star},
\label{lum_density}
\end{equation}
where $\Gamma(x) = \int^{+\infty}_{0}e^{-t}t^{x-1}dt$, and $\gamma(x,l)=\int^{l}_{0}e^{-t}t^{x-1}dt$.
To compare with previous results \citep[e.g.,][and references therein]{Sawicki:2006lr}, the faint luminosity limit 
is set as 0.1$L^{\star}$. The luminosity density at 1700{\angstrom} can be computed from 
\begin{equation}
L_{\rm 1700\rm\angstrom} = \frac{4\pi d_L^2}{1+z}10^{-\frac{2}{5}(48.6+m_R)}.
\end{equation}
The characteristic luminosity, 
$L^{\star}_{\rm 1700\rm\angstrom} = 1.2\times10^{29}$~erg~s$^{-1}$~Hz$^{-1}$ based on our best Schechter function fit.
Both the LF measured by this work and the LF measured by combining this work and the faint-end data points
from \citet{Reddy:2009kx} are used to compute the UV luminosity density. 
The total UV luminosity density derived from these two measurements are
consistent with each other, which are $2.19\pm0.08\times10^{26}$~erg~s$^{-1}$~Hz$^{-1}$~Mpc$^{-3}$
and $2.18\pm0.05\times10^{26}$~erg~s$^{-1}$~Hz$^{-1}$~Mpc$^{-3}$,
respectively.
In Table~2, we summarize the results of LF measurements and the total UV luminosity density
from this work, \citet{Reddy:2009kx}, \citet{Shim:2007uq}, and \citet{Sawicki:2006fj,Sawicki:2006lr}.
We find that our UV luminosity density results agree with that from \citet{Reddy:2009kx} within 
$1\sigma$ uncertainty, but are significantly larger than that from \citet[][by
about $6\sigma$]{Sawicki:2006fj,Sawicki:2006lr} and from \citet{Shim:2007uq}. This discrepancy is mainly 
due to the different faint-end slopes estimated. Compared to the results from \citet{Sawicki:2006fj,Sawicki:2006lr}
and  \citet{Shim:2007uq}, this work and \citet{Reddy:2009kx} suggest a much steeper faint-end slope
of the UV LF (Table~2), and that the faint LBGs make a significant contribution to the UV luminosity density.


\begin{figure}
\epsscale{1.3}
\plotone{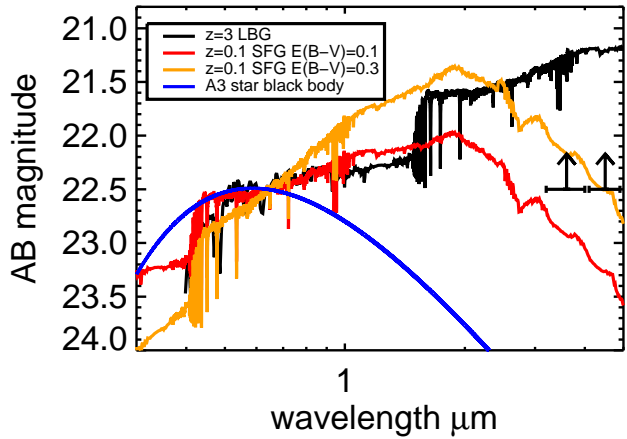}
\caption{Spectra of $z\sim3$ LBGs and low redshift interlopers: $z=0.1$
star-forming galaxies (SFG) with $E(B-V)=0.1$ and $E(B-V)= 0.3$ and an A type star. All the spectra are scaled to the $R=22.5$.
The two upward arrows represent the $5\sigma$ flux limits of the [3.6] and [4.5] band
in Bo\"otes field.\label{interloper}}
\end{figure}

\subsection{Systematic uncertainties of LF measurement}
The following effects are considered for their impact on the LF measurements, especially in term of the possibility of introducing systematic errors:
\begin{itemize}
\item Contamination fraction: for the LF measurement, we use the fraction of low redshift 
interlopers from \citet{Reddy:2008fj},
in which the spectroscopically confirmed sample is used to constrain the contamination
rate in each magnitude bin.
At the faint end, the fraction approaches zero, while it
is about 70\% at the bright end. The reasons why the contamination
rate is low at the faint end are (1) Galactic type A stars are not 
as faint as $R=25$; and (2) the LF of low redshift galaxies ($z\sim0.1$)
becomes flat at the faint end and the survey volume at $z\sim3$
is about two orders of magnitude higher than that at $z\sim0.1$.

We use  {\it Spitzer} IRAC photometry to estimate the 
bright-end contamination rate of low redshift interlopers. The majority of the interlopers are
A type stars and star-forming galaxies at $z\sim0.1$.
Figure~\ref{interloper} shows the observed-frame model SED of $z\sim3$ LBGs and those two 
types of interlopers in the wavelength range from 3000{\angstrom} to $5\mu$m.
The spectra of star-forming galaxies are generated using the same procedure as 
described in \S4.2 and then shifted to $z=0.1$ and $z=3$.
Figure~\ref{interloper} suggests that the interlopers, both A type stars and star-forming galaxies at $z=0.1$,
are expected to have bluer $R-[3.6]$ and $[3.6]-[4.5]$ colors than $z\sim3$ LBGs.
We use the $R-[3.6]$ and $[3.6]-[4.5]$ colors to estimate
the contamination rate of galaxy candidates in the $R$-band magnitude range between 22.0 and 22.5.
The magnitude limit in the [3.6] band is about 22.5 AB ($5\sigma$).
As shown in Figure~\ref{interloper}, the bright $z\sim3$ LBGs should have
firm detections in both the [3.6] and [4.5] bands \citep[e.g.,][]{Bian:2012lr}, 
however, neither A type stars nor $z=0.1$ star-forming galaxies with $E(B-V)=0.1$ 
can be detected in both the [3.6] and [4.5]
bands.  Star-forming galaxies at $z=0.1$ with $E(B-V)=0.3$ can be detected in the [3.6] band and marginally in the [4.5]
band, but they have much bluer R-[3.6] and [3.6]-[4.5] colors than those in LBGs.
There are about 360 LBG candidates in the $R$-band magnitude range between 22.0 and 22.5; 
among them, about 110 candidates have both [3.6] and [4.5] detections and $[3.6]-[4.5]>0.0$.
The latter color cut will exclude most of the galaxies/AGNs lower than $z\sim1.5$ \citep[e.g.,][]{Donley:2008fk}.
Our follow-up spectroscopic observations have shown that this color cut is very robust in rejecting 
contamination from low redshift interlopers (Bian et al. 2012 in preparation). 
This result indicates that the contamination rate of low redshift interlopers in the bright LBG candidates is indeed about $70\%$.  
We cannot distinguish AGNs/quasars from LBGs at $z\sim3$ using $[5.8]-[8.0]$ color \citep{Lacy:2004fk,Stern:2005lr,Donley:2008fk} due to the shallow [5.8] and [8.0] imaging data. 


\item Redshift distribution: 
when we calculate the luminosity of LBGs, all LBGs are assumed to be
at the same redshift ($z=2.9$) rather than in a redshift distribution.
The FWHM of the redshift distribution (Figure~\ref{completeness}) is about 0.6 ($2.7<z<3.3$),
which scatters the LBGs from a given absolute magnitude bin 
into neighboring magnitude bins.
There are more LBGs in the fainter absolute magnitude
bin, introducing a bias, especially at the bright end. 
To estimate the influence of this effect, 
we conduct a Monte Carlo (MC) simulation. We focus on the 22.0-22.5, 22.5-23.0
and 23.0-23.5 magnitude bins. Each magnitude bin is divided into 10 sub-bins.
The number of galaxies is generated for each sub-bin following the Schechter function  (equation~\ref{schechter}),
and the scatter in apparent magnitude is mainly due to the redshift distribution in Figure~\ref{completeness}.
The final number of galaxies in the 22.5-23.0 magnitude bin is compared with
the initial number of galaxies falling into this magnitude bin, and we find
that the LF for the 22.5-23.0 magnitude bin can be overestimated by $\sim18\%$ due to 
this effect. For comparison, the statistical error for this magnitude bin is $7\%$.


\item Galaxy spectral template model: 
the effective comoving volume calculation depends on the galaxy spectral template.
The spectral template used in our simulation is a spectrum
of an idealized galaxy with a  300~Myr old stellar population and a constant SFR, solar metallicity and 
Salpeter IMF, and an $E(B-V)$ distribution from \citet{Reddy:2008fj}. All these parameters
can affect the value of V$_{\rm eff}$. We perform an MC simulation to estimate the
effective comoving volume with different ages (from 100~Myr to
1~Gyr) and find that the effective comoving volumes change by less than 5\% due to the age,  
which is consistent with the results in \citet{Sawicki:2006fj}. Another factor that introduces 
uncertainty is the distribution of $E(B-V)$. We allow the fraction of the LBGs in 
each $E(B-V)$ bin to vary by $20\%$ and perform an MC simulation, and find 
that the uncertainty in the LF caused by the $E(B-V)$ variation is less than $10\%$.
The spectral template does not include the Lyman $\alpha$ emission/absorption,
which could also influence our results.
In the redshift range to which the selection criteria are sensitive, the Ly$\alpha$ line
falls into the $B_{\rm W}$-band filter. For an LBG with observed-frame Ly$\alpha$ equivalent width of 
50~\angstrom, the real intrinsic $U_{\rm spec}-B_{\rm W}$/$B_{\rm W}-R$ color will be 0.05 magnitude redder/bluer than that in our simulation,
which will make galaxies at lower redshift fall into the selection criterion, but the difference is much 
smaller than the typical uncertainty of a $U_{\rm spec}$-band magnitude ($\approx0.5$) or a 
$B_{\rm W}$-band magnitude ($\approx0.1$). This effect will influence the effective
volume estimation by less than $3\%$.
Therefore, neither the galaxy age in the template spectra nor 
the Ly$\alpha$ emission/absorption line makes a large impact on the selection function.


\item Cosmic variance: cosmic variance is 
another possible source of systematic uncertainty for the LF measurement due to the limited survey volume and 
the fluctuations of the
dark matter density on large scales. Using the cosmic variance calculator\footnote{http://casa.colorado.edu/$\sim$trenti/CosmicVariance.html} 
\citep{Trenti:2008lr}, we find that the cosmic variance for the Bo\"otes field is about $4\%$ for a minimum halo mass of 
$8\times10^{11}~M_{\sun}~h^{-1}$. For comparison, the cosmic
variance in a one deg$^2$ area is $9\%$ for a minimum halo mass of 
$8\times10^{11}~M_{\sun}~h^{-1}$.  Cosmic variance
dominates the uncertainty of the LF at the bright end compared to Poisson errors. 
\end{itemize}

We conclude that the greatest degree of uncertainty in the LF measurement
comes from the contamination
fraction; all other factors combined will contribute $\la30\%$ uncertainty to
the LF measurement. 


\subsection{Discussion}


The large survey field also allows us to 
select a sizable sample of luminous LBGs with $-23>M_{1700{\rm \angstrom}}>-25$ ($L> 6L^\star$), probing
the UV LF in this range for the first time.
Our measurement
(the brightest three points)
shows an excess power compared to the Schechter function fit. 
\citet{van-der-Burg:2010vn} found a similar excess power
in the luminosity range $-23>M_{1700{\rm \angstrom}}>-23.5$ (green triangles in Figure~\ref{lf}),
and they suggested that it is due to gravitational lensing. In Figure~\ref{lf}, we show
the $z\sim3$ quasar LF \citep{Hunt:2004fk} and find that the bright end LBG LF follows the 
quasar LF well, which suggests that the majority of the excess power
can be explained by the LF of $z\sim3$ quasars. 
We subtract the quasar LF value from these three points to statistically remove the quasar contribution.
The three red squares present the residual values, which are still higher than the
best-fit Schechter function. It is worth noting that the uncertainty of the quasar LF measurements
from \citet{Hunt:2004fk} is large due to the small size of the faint quasar sample. Therefore,
the excess power of the LBG LF can be within the uncertainty of the quasar LF measurements. 
If the excess is real, it can be caused by gravitational lensing, which boosts the fainter LBGs to the bright
end \citep[e.g.,][]{Jain:2011lr}. It is also possible that the LF of LBGs actually shows excess power at the bright
end. The similar excess power at the bright end ($L>2L^\star$) of the UV LF has been found in the $z\sim$7-8 LBG sample
\citep[e.g.,][]{Yan:2011qy,Yan:2012oq}. The bright end cutoff of the UV LF is regulated by 
feedback processes and dust obscuration \citep[e.g.,][]{Lacey:2011qy}. 
If the excess power is real, that would suggest that those physical mechanisms are probably not efficient in these
most UV luminous LBGs. 
To have an accurate measurement of the bright end LF,
follow-up spectroscopic observations for the bright LBGs
are required (Bian et al. 2012, in preparation).


\section{Near-IR LUMINOSITY FUNCTION}\label{nir_lf}
The {\it Spitzer} Deep-Field Survey \citep{Ashby:2009kq} covers the
 whole 9~deg$^2$ NDWFS Bo\"otes field with all four IRAC bands at wavelengths of 
3.6~$\mu$m, 4.5~$\mu$m, 5.8~$\mu$m, and 8.0~$\mu$m. 
There are 4 epochs in the survey with a total exposure time of
$12\times30$~s. 
In this paper,  we use the IRAC2 (4.5~$\mu$m, [4.5]) band to probe the rest-frame
near-IR emission of LBGs at $z\sim3$.
The advantages of the [4.5] band are:  (1) the depth of the [4.5] band is comparable to 
the [3.6] band depth; 
(2) the [4.5] band is less influenced by the AGN
power-law component than the [5.8] and [8.0] bands; and (3) the rest-frame wavelength
of [4.5] at $z\sim3$ is at 1.1$\mu$m ($J$-band at $z\sim0$), which  
probes the evolved stellar population and is not affected by strong emission lines
that bias the observed-frame IR continuum measurements.
As discussed in section \S5.3, the [4.5] detection rate is high ($\sim100\%$) in the bright LBGs ($R<23.0$).
The detection rate decreases quickly with increasing magnitude, and the [4.5] band detection rate for
the faintest magnitude bin ($24.5<R<25.0$) is about 
50\%. The rest-frame near-IR LF is derived based on the UV LF and the relation between the $R$-band magnitude
and the $R-[4.5]$ color in the LBGs at $z\sim3$. We follow the method developed by \citet{Shapley:2001lr},
who estimated the $K$-band (rest-frame V-band) LF of LBGs at $z\sim3$ based on their UV LF and UV-optical colors.

\begin{figure}[]
\epsscale{1.2}
\plotone{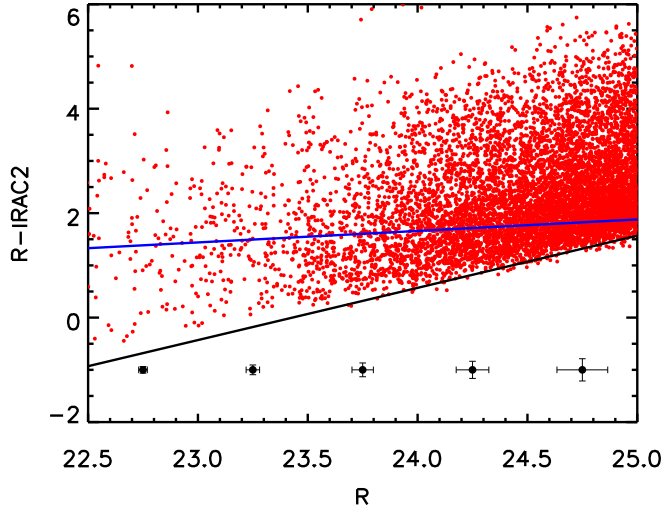}
\caption{Relation between $R$ magnitude and $R-[4.5]$ color (red filled points). The black solid
line shows the typical magnitude limits of the [4.5] band, and the blue solid line
represents the best-fitted linear regression line with survival analysis method. 
The black filled points with error bar represent
typical errors of $R$-band magnitude and $R-[4.5]$ color in the individual magnitude bins.\label{r_irac2}}
\end{figure}

\begin{figure}[ht]
\epsscale{1.25}
\plotone{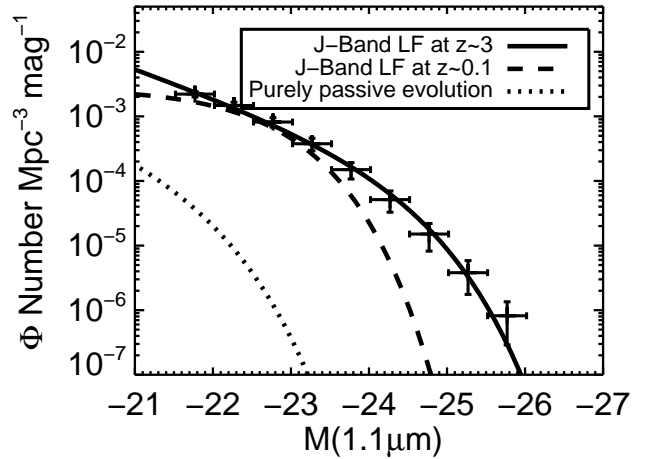}
\caption{$z\sim3$ LBG LF of [4.5] band (rest-frame $J$-band).
The solid line represents the best-fit Schechter function 
for the [4.5] band (rest-frame $1.1\mu$m) LF, and the dashed line represents the $J$-band LF for the galaxy
in the local Universe. The dotted line is the predicted $J$-band LF at $z\sim0.1$, if we assume a 
purely passive evolution in these galaxies from $z\sim3$.\label{irac2_lf}}
\end{figure}

\begin{figure*}
\epsscale{0.8}
\plotone{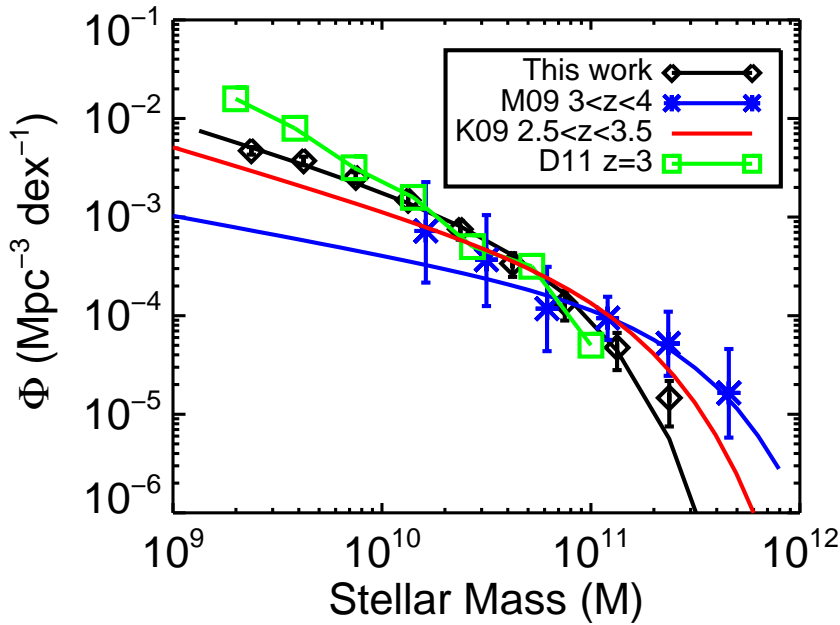}
\caption{Stellar mass functions of $z\sim 3$ LBGs from this work (black solid line), 
and K-selected galaxies 
\citep[K09, red curve, M09, blue curve]{Kajisawa:2009lr, Marchesini:2009kx}.
The green
curve represents the $z\sim3$ stellar mass function derived from 
cosmological simulations with momentum-conserved wind feedback model \citep[D11]{Dave:2011fj}.\label{SMF}}
\end{figure*}

\begin{figure*}
\centering
\epsscale{0.7}
\plottwo{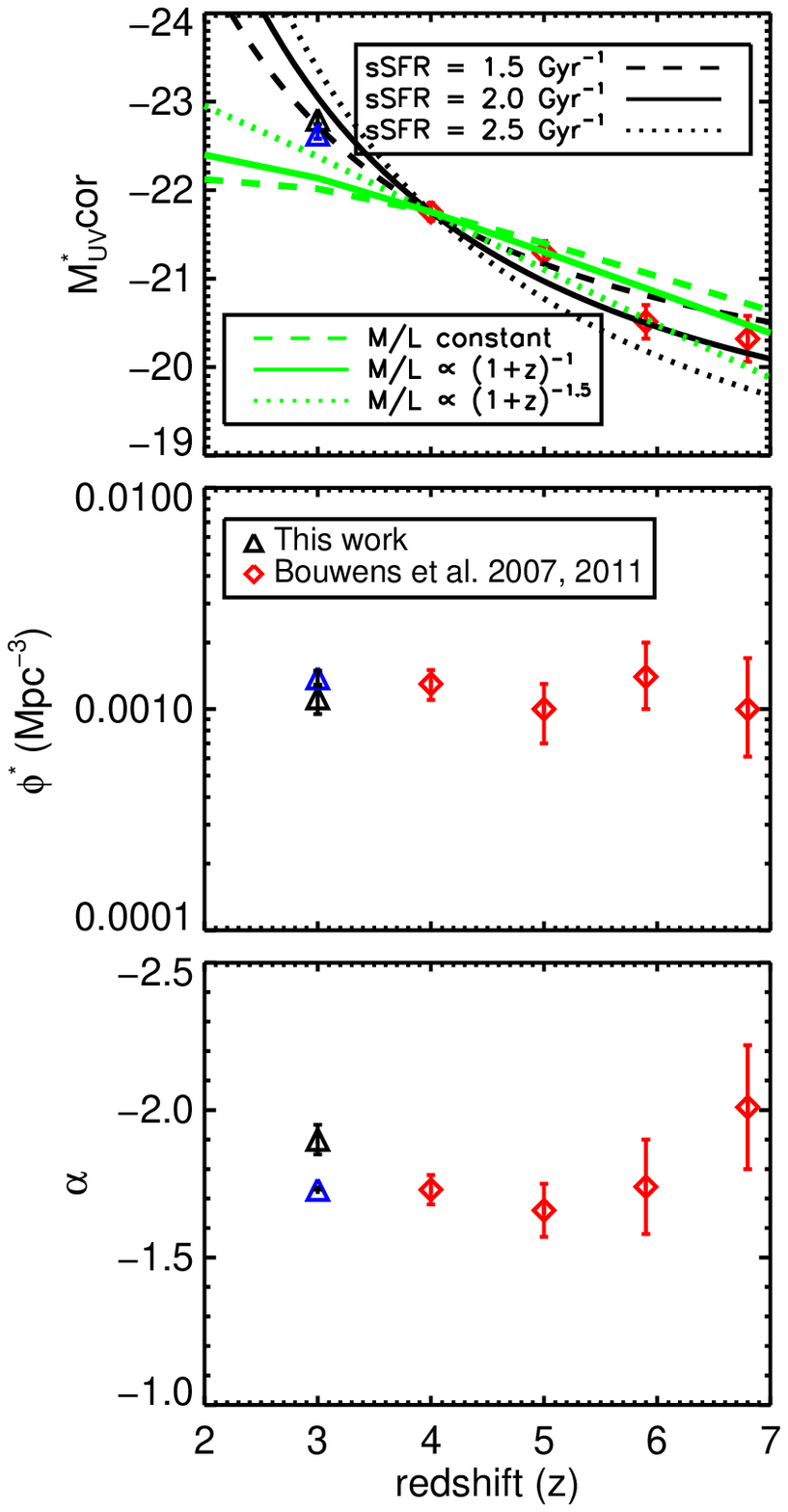}{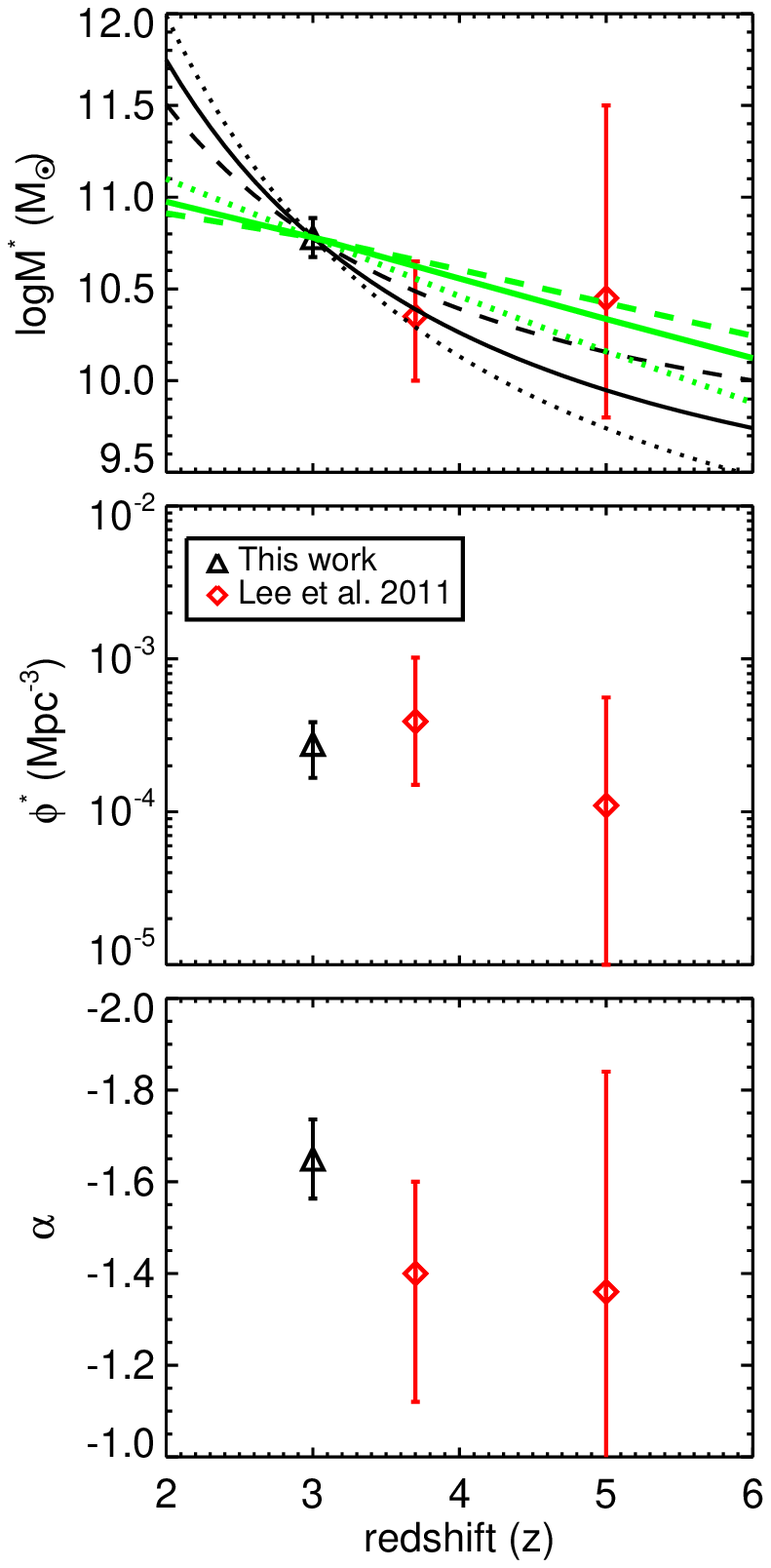}
\caption{Evolution of best-fit parameters of Schechter functions of UV LF (left panel) and SMF (right panel)
with redshifts. Left panel: For characteristic absolute magnitude, we use the dust-corrected value ( $M^{\star}_{\rm UV}$,cor).
The black triangles are derived by fitting the Schechter function with
the combination of LF from this work (bright end)
and \citet[faint end]{Reddy:2009kx}, and the blue ones are derived by fitting the Schechter function with
the LF from this work with a fix faint end slope, $\alpha=1.73$.
The red diamonds are from \citet{Bouwens:2007lr,Bouwens:2011uq}.  
Right panel: The black triangle is 
from this work, and the red diamonds are from \citet{Lee:2011fj}. 
The prediction of two different galaxy evolution models
are plotted in both panels. The green curves represent the episodic
star formation model with different assumptions on evolution of the halo mass 
to dust-corrected UV luminosity or stellar mass ratio.
The green dashed, solid and dotted curves represent models with the ratio constant with cosmic time,
$\propto(1+z)^{-1}$, and $\propto(1+z)^{-1.5}$, respectively.
The black solid curves represent the smoothly rising star formation model.
The dashed, solid and dotted black curves represent models with sSFR of 1.5~Gyr$^{-1}$,
2.0~Gyr$^{-1}$, and 2.5~Gyr$^{-1}$, respectively. 
\label{z_evol}}
\end{figure*}


\subsection{The Rest-frame UV/Near-IR Color Relation of LBGs}
We carry out  photometry on the [4.5] band images using
SExtractor with the same parameters as in Table~2 of \citet{Ashby:2009kq}, except that we use a 
lower detection threshold ($DETECT_-THRESH=1.0$) to detect fainter 
sources. We use the 4$''$ aperture magnitude as the output magnitude.
We first apply the aperture correction to the magnitude and then
covert the Vega magnitude to AB magnitude.
The  2.5~$\sigma$ limiting magnitude is 23.4 AB.
Finally, the positions of sources are matched with 
the LBG positions.

The distribution of $R$-band magnitude versus $R-[4.5]$ color is shown
in Figure~\ref{r_irac2}. The solid line represents the magnitude limit of the 
[4.5] band.
As there are only upper limits of $R-[4.5]$ color for 
a large fraction of LBGs, 
we used one of the survival analysis methods, 
the schmittbin method based on maximum likelihood 
estimation \citep{Schmitt:1985zr},
in the Astronomy Survival Analysis (ASURV) \citep{Isobe:1986mz}
in the IRAF package, stsdas.analysis.statistics, 
to take into account the upper limits of $R$-[4.5] color.
We calculated the linear regression between $R$ and $R-[4.5]$ color. 
and found that the slope, $\Delta$($R-[4.5]$)/$\Delta R$, is 0.22. 
The probability of correlation between $R$ and $R-[4.5]$ color is greater than 99.98\%.
We also use other methods in the ASURV package, e.g., emmethod and buckleyjames method,
and obtain similar results.


\subsection{Results}
We estimate the near-IR LF using simulated galaxy colors based on the UV LF measurements and the rest-frame UV/near-IR color relation, following \citet{Shapley:2001lr}: 
(1)  we generate $100,000$
galaxies with $R$-band magnitudes in the range from 22.5 to 27.0.
The distribution of the $R$-band magnitudes
follows the distribution of the UV LF in \S\ref{lfresults};
(2) we derive the distribution
of $R-[4.5]$ in the $23.4<R<23.5$ magnitude bin. In this 
magnitude bin, the [4.5] band detection rate is about 75\%, and the
quasar/AGN contamination rate is low.
We assume the $R-[4.5]$ distribution is Gaussian.
At $R\approx23.5$, the $R-[4.5]$ color distribution is truncated at $R-[4.5]$ $\approx0$ 
due to the detection limit of the [4.5] band. Therefore, 
we only use galaxies with $R-[4.5]$ color values in the range
between 0 and 4 to fit the Gaussian function;
(3) we generate $100,000$ $R-[4.5]$ colors that follow the Gaussian distribution 
function derived in step (2).
For each $R-[4.5]$ value, an offset ($\Delta$($R-[4.5]$)) is applied based on its $R$-band magnitude,
$\Delta$($R-[4.5]$) = 0.22($R-23.45$), to get the final relation between $R$-band magnitude
and [4.5] magnitude for each galaxy. Using this method, 
the [4.5] magnitudes for a sample of 100,000 galaxies are generated
based on their $R$-band magnitude and the relation between $R$ magnitude and $R-[4.5]$ color,
and the [4.5] band LF is derived; 
(4) we use a MC simulation to estimate the uncertainty of the near-IR LF.
10,000 simulated UV LFs 
are generated based on the uncertainty of the UV LF measurements in \S\ref{lfresults}.  
The same procedure as above is used to transfer the $R$-band magnitude 
to the [4.5] magnitude in $z\sim3$ LBGs. 
When we use the $R$-band magnitudes
to calculate the [4.5] magnitudes,
the uncertainties, including
the $R$-band and [4.5] band photometric errors, the uncertainty of the slope between $R$ and $R-[4.5]$ color,
and the uncertainty of the Gaussian distribution of the $R-[4.5]$ color for the given $R$ magnitude range,
are also considered. Finally,   
a total of $100,000$ simulated near-IR LFs are derived. 
The standard deviations of the 10,000 near-IR LFs in each magnitude bin are considered as the 
uncertainties of the near-IR LF; and
(5) we fit the [4.5] band (rest-frame 1.1$\mu$m) LF with a
Schechter function (Figure~\ref{irac2_lf}) and find that
the best-fit Schechter function parameters for the rest-frame 1.1$\mu$m LF of $z\sim3$ LBGs are 
$\phi^{\star}_J=(3.1\pm1.9)\times10^{-4}$~Mpc$^{-3}$, $M^{\star}_J = -24.00\pm0.30$, and
$\alpha_J = -2.00\pm0.17$.


\subsection{Discussion}

Figure~\ref{irac2_lf} compares the rest-frame 1.1$\mu$m LF of $z\sim3$ LBGs to the $J$-band LF of
nearby galaxies at $z\sim0.1$ \citep{Cole:2001fk}.
The best-fit Schechter function parameters for nearby
galaxies are $\phi^{\star}_J=(3.57\pm0.36)\times10^{-3}$~Mpc$^{-3}$, $M^{\star}_J = -23.13\pm0.02$, and
$\alpha_J = -0.93\pm0.04$. The rest-frame $J$-band LFs show significant evolution from $z=3$ to $z\sim0.1$. 
Compared to the local LF, the rest-frame $J$-band LF at $z\sim3$ has smaller  
$\phi^{\star}_J$ by an order of magnitude,  a steeper
faint-end slope ($\alpha$), and brighter characteristic luminosity ($M^{\star}_J$) by $\approx1.0$ magnitude.

These trends are also found in the rest-frame optical ($V$-band) LF between $z\sim3$ and $z\sim0$ 
\citep{Shapley:2001lr}. The steep faint-end slope of the rest-frame near-IR LF is mainly due to the steep
faint-end slope of the rest-frame UV LF and the positive correlation between $R$ magnitude and 
$R-[4.5]$ color. A similarly steep faint-end slope ($\alpha=-1.85\pm0.15$) is also found in the optical ($V$-band) LF of LBGs at $z\sim3$
\citep{Shapley:2001lr}, which is consistent within the errors with our near-IR faint-end slope,
and the optical LF in \citet{Shapley:2001lr} is derived based on a UV faint-end slope of $\alpha\sim-1.57$.
If the authors adopt a much steeper UV faint-end slope, such as that in \citet{Reddy:2009kx} and this work,
the faint-end slope of the optical LF will get even steeper. 
One of the caveats for the rest-frame near-IR LF derived in this work is that the IRAC depth of the LBG sample is shallow.
We obtain the $R-[4.5]$ color distribution based on a bright magnitude bin and assume that this distribution does
not change in the fainter magnitude bins,
therefore, it is unclear
whether the $R-[4.5]$ color distribution and the positive slope between $R$ magnitude and $R-[4.5]$ color will still
hold for the faint LBGs.

To derive the 
rest-frame $J$-band luminosity density of LBGs at $z\sim3$, we integrate the 
Schechter function to a faint luminosity limit, 0.1~$L^{\star}$, which is about the
magnitude limit of the [4.5] band images
and find $\log_{10}\rho_J$ (erg~s$^{-1}$~Hz$^{-1}$) $= 27.04^{+0.09}_{-0.11}$.

We consider a simple purely passive evolution model for the near-IR LF.
Using the BC03 model, we generate SEDs of galaxies
with a constant star formation history of $300$~Myr, which is about the typical age of LBGs
at $z\sim3$ \citep{Shapley:2001lr}, then we shut down the star formation in the galaxies and make
these galaxies passively evolve for the following 11.0~Gyr, from $z\sim3$ to $z\sim0.1$. 
We find that the absolute $J$-band magnitude (after reddening with
$E(B-V)=0.15$) at $z\sim3$ right after shutting down star formation is about 2.7 magnitudes brighter 
than that at $z\sim0$. In Figure~\ref{irac2_lf}, the dotted curve represents the predicted local near-IR LF 
based on this purely passive evolution model. 

\subsection{Stellar mass function}
Since the near-IR is a good tracer of the old and evolved stellar population, 
the rest-frame near-IR LF can be used
to derive the SMF of LBGs at $z\sim3$. 
We adopt the relation between the rest-frame $1.1\mu$m absolute magnitude and the stellar mass
in \citet{Reddy:2012qy}:
\begin{equation}
\log(M_{\rm stellar}/M_{\sun})=-(0.56\pm0.09)M_{1.1}-(2.42\pm1.94),
\end{equation}

Using this relation,  the faint-end slope of the SMF will be flatter compared to that of the rest-frame $J$-band LF.
In Figure~\ref{SMF}, we show the SMF and best-fit Schechter function of LBGs at $z\sim3$.
The best-fit parameters are $\phi^{\star} = (2.8\pm1.1)\times10^{-4}$~Mpc$^{-3}$, $\log_{10}M^{\star}_{\rm stellar} (M_{\sun})=10.78\pm0.11$,
and $\alpha=-1.65\pm0.09$. The large survey area allows us to reduce the cosmic variance,
which contributes significant uncertainty to the previous SMF measurements \citep[e.g.][]{Marchesini:2009kx}.
In Figure~\ref{SMF}, we also show the SMFs derived from $z\sim3$ $K$-selected galaxy samples
in deep field surveys \citep{Marchesini:2009kx,Kajisawa:2009lr} and the galaxy SMF at $z\sim3$ predicted
by cosmological hydrodynamic simulations \citep{Dave:2011fj}.
For comparison, we scale the galaxy stellar mass derived based on different IMFs to 
the mass based on a Kroupa IMF \citep{Kroupa:2001vn}. 
The $K$-selected galaxy sample in \citet{Marchesini:2009kx} is selected from about 400~arcmin$^2$,
which is about 4 times larger than that in \citet{Kajisawa:2009lr}; on the other hand, the survey depth in \citet{Kajisawa:2009lr}
is about 1-2 magnitudes deeper than that in \citet{Marchesini:2009kx}. Therefore, \citet{Marchesini:2009kx}
put stronger constraints on the SMF at the high mass end, while \citet{Kajisawa:2009lr} measure the 
low mass end of the SMF more accurately. 

At the high mass end, $M^{\star}_{\rm stellar}$ in the LBG sample is smaller than 
those in the $K$-selected galaxy samples \citep{Marchesini:2009kx,Kajisawa:2009lr} 
at a 2$\sigma$ significance level, indicating a lower characteristic mass in the LBGs.
For galaxies with stellar mass greater than $10^{11}$M$_{\sun}$, 
the density of LBGs is significantly smaller than that of mass selected galaxies,
especially for the sample from \citet{Marchesini:2009kx}. We find that the ratio of stellar mass density
between this work and \citet{Marchesini:2009kx}/\citet{Kajisawa:2009lr} is 
$0.26^{+0.20}_{-0.15}$/$0.38^{+0.62}_{-0.26}$ in the stellar mass range
$10^{11}$--$10^{12}$~M$_{\sun}$, which suggests that LBGs are not the dominant
galaxy population at the high mass end of the galaxy SMF at $z\sim3$. 
LBGs are selected based on rest-frame UV colors, therefore, this 
method cannot select highly obscured galaxies or 
passively evolving early type galaxies. \citet{van-Dokkum:2006lr} study
a sample of massive $K$-selected galaxies ($M_{\rm stellar}>10^{11}M_{\sun}$) 
and find that the LBGs make up only 20\% of the sample, and the rest of the sample
are distant red galaxies \citep{Franx:2003fk}.


At the low mass end, the SMF shows a somewhat steeper slope, although this difference is statistically insignificant. The value of the slope
agrees with that in \citet{Marchesini:2009kx}
($\alpha=-1.39^{+0.63}_{-0.55}$) and \citet{Kajisawa:2009lr} ($\alpha=-1.63^{+0.14}_{-0.15}$)  within 
$1\sigma$ error. This 
 suggests that the LBG population dominates the low mass end of the galaxy SMF at $z\sim3$.
This steep slope also suggests that the UV-selected star-forming galaxies make significant contributions 
to the total stellar mass density \citep[e.g.,][]{Kajisawa:2009lr,Reddy:2009kx}.
By integrating the LBG SMF, we find the stellar mass density of $z\sim3$
LBGs with stellar mass between $10^8$-$10^{10}$~M$_{\sun}$  to be $51\pm4$\% of the  total $z\sim3$ LBG stellar mass density,  which agrees with the result in \citet{Reddy:2009kx}.

In the mass range between $10^{9.5}$ and $10^{11}$~M$_{\sun}$,  the SMF of LBGs is roughly consistent with 
that derived from cosmological hydrodynamic simulations \citep{Dave:2011fj}.  In this type of simulation,
stellar mass is regulated by momentum-driven winds \citep{Murray:2005lr} and recycled
wind mode accretion \citep{Oppenheimer:2010fk}. At the low mass end, the model overproduces
the number of the galaxies, showing a steeper slope with $\alpha=-2.0$.
On the other hand, the SMF at the massive end predicted by the model is consistent with that of LBGs, 
but is smaller than that in mass selected galaxies; this may reflect the finite simulation volume that under-predicts the massive galaxy population.

\section{The Evolution of the UV LF and SMF}\label{evol}
We compare our $z\sim3$ UV LF and SMF results with those
 from other high redshift LBG ($z>4$) studies
\citep{Bouwens:2007lr,Lee:2012lr} to study the evolution of the UV LF and SMF with cosmic time.
Figure~\ref{z_evol} shows 
how the best-fit Schechter function parameters, including dust-corrected $M^{\star}_{\rm UV}$/$M^{\star}_{\rm stellar}$, 
$\phi^{\star}$, and $\alpha$ in UV LF and SMF,  evolve with redshift.
For the UV LF evolution, the black open triangle at $z\sim3$ is from this work.
The LF measurements for LBGs at $z\sim4$, $z\sim5$, and $z\sim6$ are from \citet{Bouwens:2007lr},
and the data point at $z\sim7$ is from \citet{Bouwens:2011uq}. 
In this plot, we use the dust-corrected $M^{\star}_{\rm UV}$ ($M^{\star}_{UV}$,cor) rather than observed $M^{\star}_{\rm UV}$,
as the dust-corrected  $M^{\star}_{\rm UV}$ can be used to represent the SFRs in the galaxies.
\citet{Bouwens:2009uq,Bouwens:2012fk} measure the UV-continuum slope, which is a direct indicator of the dust extinction 
 in LBGs from $z\sim3$ to $z\sim7$.  We adopt their UV slope measurements and the relation between
 the UV-continuum slope ($\beta$) and the UV dust extinction ($A_{\rm UV}$)
 \citep{Meurer:1999fk} to correct $M^{\star}_{\rm UV}$ at different redshifts. 
For the evolution plot of the SMF, the black triangle is from this work and the two red diamonds
at $z\sim4$ and $z\sim5$ are from \citet{Lee:2012lr}. 
In both the UV LF and SMF, the parameters, $\alpha$ and $\phi^{\star}$,
are roughly constant with redshift.  
On the other hand, the dust-corrected characteristic luminosity, $M^{\star}_{\rm UV}$,cor, in the UV LF  
increases with cosmic time.
The uncertainty in the characteristic mass, $M^{\star}_{\rm stellar}$, in the SMF
is large in this case, thus it is hard to tell whether or not $M^{\star}_{\rm stellar}$ increases with cosmic time.
The evolutionary trend provides crucial information on how galaxies built up
their mass in the early Universe. Here we will use two simple toy models to interpret the evolution of the UV LF.

In the first model, we assume that the increasing $M^{\star}_{\rm UV}$,cor/$M^{\star}_{\rm stellar}$ is mainly
due to episodic star formation through mergers.
In this model, small dark matter halos merge into larger systems
and star formation in high-z galaxies is episodic with a duty cycle of $\sim25\%$. 
The increasing $M^{\star}_{\rm UV}$/$M^{\star}_{\rm stellar}$ mainly
reflects the increasing mass of host dark matter halos with cosmic time.
\citet{Bouwens:2007lr,Bouwens:2008fk} use this model to interpret the evolution of $M^{\star}_{\rm UV}$ in 
the UV LF from $z\sim6$ to $z\sim4$. 
We follow the method in \citet{Bouwens:2007lr} to determine how the halo mass near the knee of the UV LF/SMF increases with time based on  the halo mass function of
\citet{Sheth:1999fk}. We assume three different halo mass to UV luminosity/stellar mass ratio relations:
(1) the halo mass to light/stellar mass ratio is constant with cosmic time \citep[green dashed curve -][]{Bouwens:2007lr};
(2) the halo mass to light/stellar  mass ratio varies as $(1+z)^{-1}$ \citep[green solid curve -][]{Bouwens:2008fk};
and (3) the halo mass to light/stellar mass ratio varies as $(1+z)^{-1.5}$ \citep[green dotted curve]{Stark:2007fj}.
The curves are scaled by the $M^{\star}_{\rm UV}$,cor data point at $z\sim3$ in the UV LF and by the $M^{\star}_{\rm stellar}$
data points at $z\sim4$ in the SMF. The $M^{\star}_{\rm UV}$,cor/$M^{\star}_{\rm stellar}$ in relation (1)
increases with time faster than that in relation (2). 

In the second model, 
we assume that (1)  $L^{\star}$/$M^{\star}_{\rm stellar}$ galaxies are on the main sequence 
for star-forming galaxies \citep{Daddi:2007qy}, and 
the specific star formation rate (sSFR=SFR/stellar mass)
in LBGs from $z\sim7$ to $z\sim3$ is a constant and about $1.5-2.5$~Gyr$^{-1}$ \citep[e.g.][]{Stark:2009lr,Gonzalez:2010fk,Rodighiero:2011fk};
and (2) the LBGs are in a continuous growth stage at this epoch.
This model suggests a smoothly rising SFR in the LBGs \citep[e.g.][]{Finlator:2011qy,Papovich:2011uq,Lee:2011fj}. 
The long lasting star formation could be the consequence of cold mode accretion \citep[e.g.,][]{Dekel:2009lr}.
In this scenario, the LBGs duty cycle is high ($\approx1$), and $\phi^\star$
does not change with redshift. 
Therefore, the higher redshift
$L^{\star}_{\rm UV}$/$M^{\star}_{\rm stellar}$ galaxies are the progenitors of the 
$L^{\star}_{\rm UV}$/$M^{\star}_{\rm stellar}$ galaxies at lower redshift.
As the mass is being built up, $M_{\rm stellar}^{\star}$ will increase with 
cosmic time following the relation:
$M_{\rm stellar}^{\star}(z_2)$ = $M_{\rm stellar}^{\star}(z_1)\times\exp({\rm sSFR}\times(t(z_2)-t(z_1)))$. 
$L^{*}_{\rm UV}$ will increase with the same relation.
In Figure~\ref{z_evol}, the black curves are the predicted evolution of $M^{\star}_{\rm UV}$/$M^{\star}_{\rm stellar}$ 
based on the smoothly rising star formation rate history model.
The dotted, solid, and dashed curves represent different values of the sSFR,
which are 1.5 Gyr$^{-1}$, 2.0 Gyr$^{-1}$, and 2.5 Gyr$^{-1}$, respectively. 
The growth rate of $M^{\star}_{\rm UV}$/$M^{\star}_{\rm stellar}$ increases
with increasing sSFR. 
In both models, both $\phi^\star$ and $\alpha$ are expected to be constant.

In Figure~\ref{z_evol}, we compare these two models with our observations.
For the evolution of dust-corrected $M^{\star}_{\rm UV}$, 
the episodic galaxy merger / star formation history model (green curves) 
is consistent with the data points between $z\sim4$ and $z\sim7$ \citep{Bouwens:2008fk}, while
the predicted $M^{\star}_{\rm UV}$,cor is significantly smaller than the observed 
$M^{\star}_{\rm UV}$,cor at $z\sim3$. 
On the other hand, the smoothly rising star formation history model
(the black curves) can fit the evolution of $M^{\star}_{\rm UV}$,cor
from $z\sim3$ to $z\sim7$ very well. 
For the evolution of the SMF, the predicted $M^{\star}_{\rm stellar}$ in 
the episodic star formation model evolves more slowly than that in 
the smoothly rising star formation history model.
The three Schechter parameters are correlated, and the steeper faint 
end slope in our LF would result in a higher characteristic luminosity at $z\sim3$. 
Therefore, we use the Schechter function
with a fixed faint end slope, $\alpha=1.73$, to fit our LF. The dust-corrected best-fitting characteristic luminosity
is shown in Figure~\ref{z_evol} (blue triangles). This new LF fitting does not change the characteristic luminosity
significantly or our conclusion.

Due to the large uncertainties in the SMF measurements 
for high-redshift galaxies, especially for galaxies at $z\sim5$, 
comparison with models is not conclusive.
Larger surveys with smaller statistical errors and cosmic variance are needed to differentiate different models. 

Generally speaking, the ratio of UV luminosity to stellar mass,  $M^{\star}_{\rm UV}$,cor/$M^{\star}_{\rm stellar}$, 
evolves much more rapidly in the smoothly rising star formation model
than in the episodic star formation model.
This is due to the fact that the dark matter halo mass growth rate 
decreases with cosmic time,
while the SFR in galaxies increases with cosmic time from $z\sim7$ to $z\sim3$.
The halo merger rate is about 1~Gyr$^{-1}$ at $z\sim5$ for a halo mass of $10^{10}$~$M_{\sun}$
and becomes 0.3~Gyr$^{-1}$ at $z\sim3$, while the measured sSFR is about 1.5-2.5~Gyr$^{-1}$, which is about 
a factor of 2 to 10 larger than the merger rate from $z\sim5$ to $z\sim3$.
This suggests that the mass build-up in high redshift galaxies cannot be only from halo merger
processes. Other processes, i.e., cold flow accretion, must make a significant contribution to the mass
build-up process, especially in the redshift range from $z\sim5$ to $z\sim3$. Beyond $z\sim6$, the merger rate 
becomes comparable to the sSFR. So at very early epochs ($z>6$), episodic
galaxy assembly could be the dominant process responsible for building up the stellar mass of galaxies.

It is still controversial whether high redshift LBGs have continuous star formation activity from $z\sim7$ to $z\sim3$ \citep{Finlator:2011qy,Dunlop:2013kx}, 
or whether the star formation history in LBGs is much shorter with a typical timescale of $\sim300$~Myr \citep{Stark:2009lr,Lee:2011fj}. 
The relatively young stellar population \citep{Stark:2009lr} and short star formation duty cycles from clustering measurements
\citep{Lee:2009qy}
in high-z LBGs support the latter scenario, in which the star formation history is episodic. On the other 
hand, \citet{Finlator:2011qy} argue 
that the short duty cycle from LBG clustering measurements is due to 
outflow feedback suppression of star formation in low-mass halos, which results in
only a small fraction of dark matter halos (0.2-0.4) hosting galaxies. 
In this scenario,
the actual star formation duty cycle is about unity and the star formation history in LBGs is smoothly
rising.  In summary, our results on the evolution of the UV LF and SMF favor the continuous star formation history model.

\section{CLUSTERING PROPERTIES OF LBGS}\label{clusteringsection}

\begin{figure*}
\epsscale{0.75}
\plotone{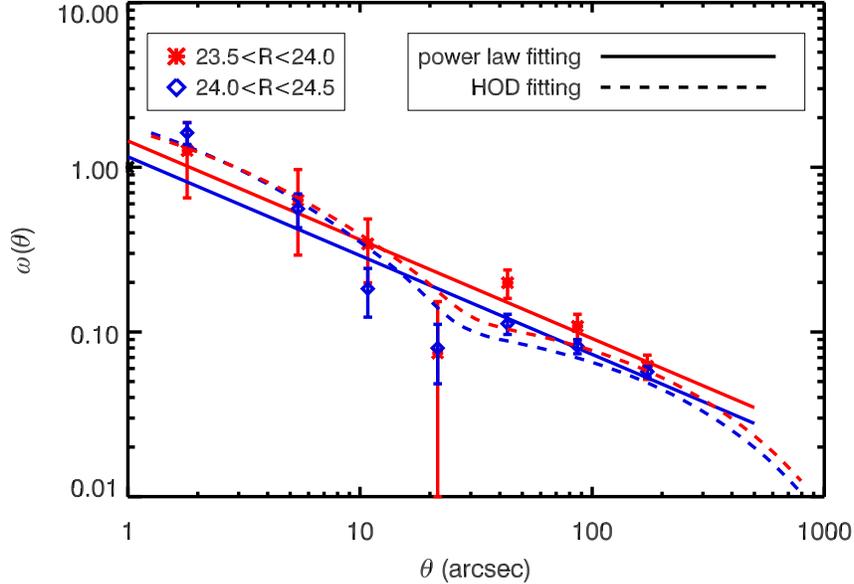}
\caption{Angular correlation functions for two subsamples at $23.5< R<24.0$ (red asterisks) and at $24.0<R<24.5$ (blue diamonds).
The red and blue solid lines present the best-fit power law of the bright and faint subsamples,
respectively. The red and blue dashed curves are the best-fit HOD models of the bright and faint, respectively.
\label{acffigure}}
\end{figure*}


In this section, we measure clustering of the bright LBGs ($L>L^{\star}$) at $z\sim3$.
The bright LBG sample is divided into two sub-samples with
 $24.0<R<24.5$ and $23.0<R<24.0$ to study the relation between clustering and LBG luminosity.
The clustering properties also allow
us to estimate the mass of dark matter halos  hosting
the LBGs. The real-space 3D correlation function can be inferred 
from the 2D angular correlation function (ACF) by using the redshift
distribution information and the Limber function \citep{Limber:1953fk, Peebles:1980kx}. 
We do not have the spectroscopic redshifts of individual galaxies,
and hence we measure the ACF and model it as $\omega=A_{\omega}\theta^{-\beta}$.
Combining the redshift distribution information obtained from our simulation (in section 5.1),
 we can obtain the clustering properties, i.e., the comoving correlation lengths for
these two sub-samples from the ACF. 

\subsection{Result}\label{subsection:acf}

We use the \citet{Landy:1993yq} estimators to measure the ACF:
\begin{equation}
\omega(\theta)=\frac{\rm{DD-2DR+RR}}{\rm{RR}},
\end{equation}
where DD, RR, and DR are the numbers of galaxy pairs with a
separation between $\theta$ and $\theta+\delta\theta$ in the
observed galaxy catalog, the random catalog, and between the observed galaxy and random catalog, respectively.
The distribution of the objects in the random catalog for individual fields
has exactly the same geometry as that in the galaxy catalog. The objects
falling into the areas with U-band coverage 
less than 720s or without $B_{\rm W}$ or $R$-band coverage have been masked
out. The number of objects in the random catalog ($n_R$)
is $\sim100$ times larger than the number of the observed galaxies ($n_G$).


The Poissonian errors for the ACF are estimated as:
\begin{equation}
\Delta\omega=\sqrt{\frac{1+\omega(\theta)}{\rm {DD}}}.
\end{equation}

We also calculate the jack-knife errors for the ACF by splitting the whole Bo\"otes field 
into 63 individual LBC fields, and we find that the jack-knife errors
are consistent with the Poissonian errors for the angle separation 
range from $0.5-200''$. This is in agreement with the result of \citet{Bielby:2012uq}.
Therefore, we adopt the Poissonian errors in the following analysis.

The $\omega(\theta)$ is calculated in each individual LBC pointing field. The 
final result is the average of the individual $\omega(\theta)$. 

The finite survey area makes the clustering results artificially
weak. The difference between the true correlation function,
$\omega_{\rm true}(\theta)$, and the measured correlation function,
$\omega_{\rm measure}(\theta)$ is a constant, which is also 
known as the integral constraint, IC \citep{Adelberger:2005fk}:
\begin{equation}
\omega_{\rm true}(\theta) = \omega_{\rm measure}(\theta)+\rm{IC}.
\end{equation}
The integral constraint is equal to the matter variance in the size
of the survey volume.
There are two approaches for estimating
the integral constraint. One of the methods is to integrate $\omega_{\rm true}(\theta)$
over the field of view \citep[see details in][]{Roche:1999lr}:
\begin{equation}
IC= \frac{\sum_i RR\omega(\theta_i)}{\sum_i RR},
\end{equation}
and the other method is to use linear theory to estimate
the galaxy variance in the survey volume. 
In this paper, we use the theoretical
estimate method to estimate the value of IC \citep{Adelberger:2005fk} (see details in Appendix \ref{ic}). 
We find the value of $IC \approx 0.02$ for both samples and add it to the measured clustering
results,  $\omega_{\rm measure}(\theta)$, to compute the values of  $\omega_{\rm true}(\theta)$.

The ACF is modeled as a power-law form: 
\begin{equation}
\omega(\theta)=A_{\omega}\theta^{-\beta},\label{acfequation}
\end{equation}
with fixed power-law index $\beta=0.6$,
which is consistent with the results in \citet{Adelberger:2005fk,Lee:2006qy}. 

The best-fit parameters for the $23.0<R<24.0$ sample are \{$A_\omega, \beta$\} =  \{$1.44\pm0.14$~arcsec$^{\beta}$, $0.60$\}
and for the $24.0<R<24.5$ sample \{$A_\omega, \beta$\} =  \{$1.13\pm0.06$~arcsec$^{\beta}$, $0.60$\}  (Figure~\ref{acffigure}).

Then, the 2D ACF is transformed into the 3D real space correlation function:
\begin{equation}
\xi = \left(\frac{r}{r_0}\right)^{-\gamma}.
\end{equation}
The parameters $r_0$ and $\gamma$ can be computed from the following relations:
\begin{equation}
A = r_0^{\gamma} B[1/2,(\gamma-1)/2]\int^{\infty}_0dzN^2r(z)^{1-\gamma}g(z)\left[\int^{\infty}_0dzN(z)\right]^{-2},
\end{equation}
\citep[see, e.g., ][]{Adelberger:2005fk,Lee:2006qy} 
\begin{eqnarray}
&&\gamma \equiv \beta+1 \nonumber\\
&&B[1/2,(\gamma-1)/2] \equiv\frac{\sqrt{\pi}\Gamma[(\gamma-1)/2]}{\Gamma(\gamma/2)}\nonumber\\
&&g(z)\equiv\frac{H(z)}{c}\nonumber \\
&&r(z) = \int^z_0 \frac{cdz}{H(z)} \nonumber \\
&&H(z) = H_0\sqrt{\Omega_{\Lambda}+\Omega_{\rm M}(1+z)^3}. \nonumber
\end{eqnarray}
From the above equations, we find that the power-law index $\gamma = 1.6$ and the comoving 
correlation lengths $r_0$ for the $23.0<R<24.0$ and $24.0<R<24.5$ LBG samples are $5.77\pm0.36 h^{-1}$~Mpc
and $5.14\pm0.16~h^{-1}$~Mpc, respectively. The comoving correlation length of the brighter LBG sample
is larger than that of the fainter LBG sample at $1\sigma$ significance level.



The LBGs in this study are about 1-2 magnitudes brighter than those in previous studies. 
But the survey area and the sample size of the bright LBGs ($R<24.5$) are
an order of magnitude larger, 
which results in better constraints on the clustering of bright LBGs. 
In our two bright LBG subsamples, 
we find that the clustering of LBGs increases with increasing galaxy UV luminosity.
This trend has been noted in previous studies of LBGs at different redshifts
\citep{Giavalisco:1998uq,Foucaud:2003lr, Adelberger:2005fk,Lee:2006qy,Hildebrandt:2007rt,Hildebrandt:2009qy}.

\citet{Lee:2006qy} and \citet{Ouchi:2005vn} found significant excess power at the small scale of the ACF ($\theta<10''$)
compared to the power-law fit in faint $z\sim3$ and $z\sim4$ LBG samples.
The excess power is mainly due to the excess number of close galaxy pairs in the same dark matter halos, e.g., the 1-halo term.
In contrast, we do not find significant excess power at the small scale 
in our bright LBG sample. This suggests that the 1-halo term contribution to the ACF decreases with increasing UV luminosity,
and satellite galaxies in dark matter halos are more likely to be faint LBGs rather than bright LBGs.

The correlation lengths in our two subsamples of bright LBGs are significantly
larger than those in fainter LBG samples from \citet{Adelberger:2005fk} and \citet{Bielby:2011qy},
in which $r_0\simeq4.0~h^{-1}$~Mpc. A similar trend was also found by
\citet{Foucaud:2003lr,Lee:2006qy,Hildebrandt:2007rt},
who also extended the clustering measurements to bright $z\sim3$ LBGs,

We find that the ACF best-fit parameters in \citet{Lee:2006qy} are similar to those in this study
in the similar UV luminosity range. However, our correlation lengths are relatively smaller than 
those in \citet{Lee:2006qy} (Table~\ref{clustering}), while \citet{Hildebrandt:2007rt}
found similar correlation lengths to those in this work (Table~\ref{clustering}). 
The main reason for this discrepancy is
the redshift distribution. \citet{Hildebrandt:2009qy} found that for LBGs at $z\sim3$ with 
$r<24.5$ the correlation lengths could vary from 5.0 to 6.0~$h^{-1}$~Mpc based on different
redshift distributions derived by various methods.
The redshift distribution of $z\sim3$ LBGs in \citet{Lee:2006qy}, 
spanning $z\sim2$ to $z\sim4$ \citep[see Figure~1 in][]{Lee:2006qy},
is much broader than that in this study, which can result in a larger correlation length. 
The redshift distributions from both works are based on simulations, and must be 
verified by spectroscopic
observations.


To obtain the mass of the dark matter halos hosting the bright LBGs, we adopt the halo occupation distribution (HOD) models
from \citet{Hamana:2004lr} and \citet{Lee:2006qy} and fit them with our ACF measurements (see details in appendix~\ref{hod}). 
The best-fit  HOD models are shown in Figure~\ref{acffigure}.
From the fitting, we can derive the minimum mass of a host halo for the galaxy population, $M_{\rm min}$, 
the typical mass of a halo hosting one galaxy, $M_1$, and the power-law
index, $\alpha$, in Equation \ref{equation:hod},
and the best-fit parameters are \{$M_{\rm min}$, $\alpha$,
$M_1$\} = \{$8.1^{+1.4}_{-1.1}\times10^{11}~h^{-1}M_{\sun}$, 0.5, $4.8\pm1.0\times10^{13}~h^{-1}M_{\sun}$\} and 
\{$(1.2\pm0.3)\times10^{12}~h^{-1}M_{\sun}$, 0.5, $7.5^{+3.0}_{-2.5}\times10^{13}~h^{-1}M_{\sun}$\} for the sample of 
LBGs with $24.0<R<24.5$ and $23.5<R<24.0$, respectively. The mean mass of the 
host halo can be estimated from:
\begin{equation}
\langle M_{\rm host}\rangle = \frac{\int^{\infty}_{M_{\rm min}}dM~M~N_g(M)~n_{\rm halo}(M)}
{\int^{\infty}_{M_{\rm min}}dM~N_g(M)~n_{\rm halo}(M)},
\end{equation}
where $n_{\rm halo}$ is the dark matter halo mass function. 
The mean masses of the hosting halos for the $24.0<R<24.5$ and $23.5<R<24.0$ samples are
 $2.5\pm0.3\times10^{12}~h^{-1}M_{\sun}$ and 
$3.3^{+0.6}_{-0.4}\times10^{12}~h^{-1}M_{\sun}$, respectively.
This result confirms that more massive dark matter halos typically
host more luminous LBGs, and are consistent with the relation between the UV luminosity and the dark matter 
halo mass being $L_{\rm UV}\propto M_{\rm halo}^{1.5}$.
This relation is similar to those for LBGs at $z\sim4$ and $z\sim5$ \citep{Ouchi:2005vn,Lee:2006qy}.


\begin{figure*}
\epsscale{0.8}
\plotone{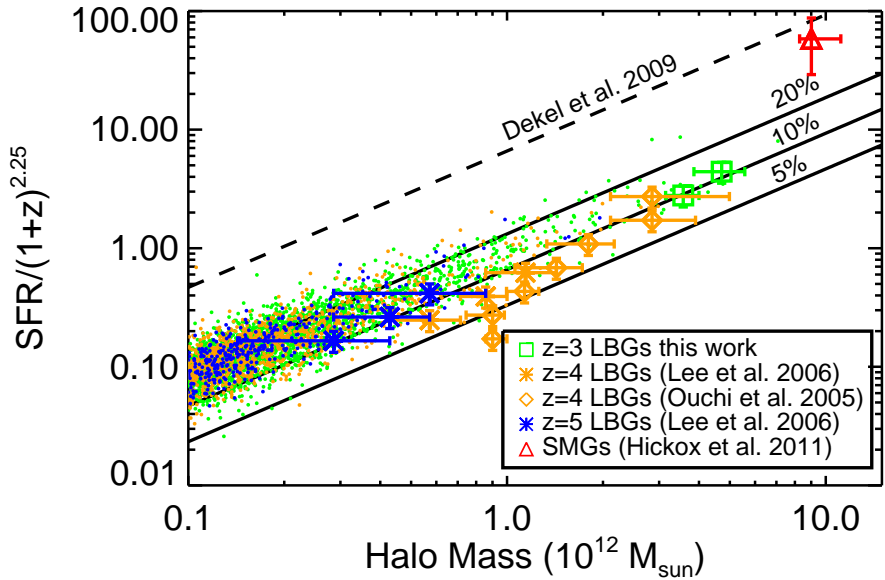}
\caption{Scaled star formation rate (SFR) versus hosting halo mass. 
The dashed line represents the relation in Eq. \ref{coldflow} derived from the cold flow model \citep{Dekel:2009lr},
and the solid lines represent the cases where the star formation efficiencies are 
20\%, 10\% and 5\% of the total cold flow accretion mass.
The LBGs at $z\sim3$ (green open squares, this work),
$z\sim4$ \citep[orange asterisks,][]{Lee:2006qy} and \citep[orange diamonds,][]{Ouchi:2005vn},  and $z\sim5$ \citep[blue asterisks,][]{Lee:2006qy}
follow the 10\% cosmic star formation efficiency line very well, while the SMGs (red triangle) at $z\sim2$ are about 0.85 dex higher.
The filled circles are the predictions of the cosmic star formation efficiency from the cosmological simulations with momentum-driven gas outflows recipe \citep{Oppenheimer:2008lr}. The points are color coded by redshift (green z=3, orange z=4, and blue z=5).
\label{fig:coldflow}} 
\end{figure*}

\subsection{Discussion}
We find that LBGs with higher UV luminosities have greater clustering strengths
and live in more massive dark
matter halos. $R$-band magnitude, which corresponds
to the rest-frame UV brightness at redshift $z\sim3$, is a good tracer for the
unobscured star formation in the galaxies. Therefore, this trend also suggests
that galaxies with higher SFRs are in more massive 
dark matter halos. This correlation can be understood in the context of the cold
flow mode of galaxy formation: gas is accreted onto dark matter halos
from the IGM and finally falls into galaxies within the dark matter halos.
This process provides the material to form stars in galaxies.
\citet{Dekel:2009lr} derive the corresponding baryonic growth rate
in the halo by fitting the growth rate of dark matter halos within the 
framework of the Extended Press-Schechter (EPS) formalism \citep{Neistein:2008lr}.
They find that the baryonic accretion rate ($\dot{M}$)
is a function of halo mass ($M_{\rm halo}$) and redshift ($z$):
\begin{equation}
\dot{M}=6.6 \left(\frac{M_{\rm halo}}{10^{12}M_{\sun}}\right)^{1.15}(1+z)^{2.25}\frac{f_b}{0.165}~M_{\sun}~\rm{yr}^{-1},
\label{coldflow}
\end{equation}
where $f_b=0.165$ is the baryonic fraction in the halos. 
This relation shows that the cold mode accretion rate increases with increasing redshift
and halo mass, suggesting that baryonic accretion is an important process to feed the star formation
in high-z star-forming galaxies \citep{Dekel:2009lr}.
This model describes how gas is accreted onto dark matter halos at large scales, but does not include any physical
processes down to the galaxy scale on how to convert the accreted gas into stars 
and to regulate star formation. Our measurements of the SFRs in the galaxies and
their host dark matter masses will allow us to connect the large scale baryonic accretion process
and the small scale star formation process.

Figure~\ref{fig:coldflow} shows the redshift-scaled SFR (SFR/$(1+z)^{2.25}$) as a function of the host halo mass.
The SFRs are estimated from dust-corrected UV luminosities.
They are derived from the $R$-band magnitude for $z\sim3$ LBGs in this work, $z$-band magnitude
for $z\sim4$ and 5 LBGs in \citet{Lee:2006qy}  and $i$-band magnitude for $z\sim4$ LBGs in \citet{Ouchi:2005vn},
by assuming a typical dust extinction $E(B-V)=0.15$ for LBG samples at $z\sim3$ and $z\sim4$
and $E(B-V)=0.10$ for the LBG sample at $z\sim5$ \citep[e.g.,][]{Bouwens:2009uq}. 
In Figure~\ref{fig:coldflow}, the dashed line represents 
the relation of 
Equation \ref{coldflow}, and the three solid lines from top to bottom
represent 20\%,  10\%, and 5\% of accreted baryonic gas converted into star
formation. 


From Figure~\ref{fig:coldflow}, we find that the observed redshift-scaled 
SFR as a function of mass from this and other 
work \citep{Ouchi:2005vn,Lee:2006qy} follows the trend 
predicted by the baryonic accretion model very well over about two orders of magnitude
of dark matter halo mass. 
This relation suggests that the star formation in LBGs is fueled 
by these baryonic flows and that the cosmic star formation efficiency ($\rm sSFE=\rm SFR/\dot(M)$),
which is defined as the efficiency of the conversion of cold flow accretion gas into star formation, is 
between $5\%$ and $20\%$. The cosmic star formation efficiency does not change significantly
with redshift ($3<z<5$), dark halo mass ($10^{11}-10^{13}~M_{\sun}$), 
or LBG luminosity ($0.1L^*-3L^*$).
The low efficiency is consistent with cosmological simulations 
\citep[e.g.,][]{Oppenheimer:2008lr} in which the cosmic star formation efficiency is about $20\%$ (Figure~\ref{fig:coldflow}).

There are two possibilities for the relatively constant cosmic star formation efficiency in a large
range of dark matter halo mass. (1) The cosmic star formation efficiency may be set by momentum-driven outflows, which can eject up to 80\% of the total inflow at the smallest halo masses shown, and by retardation of accretion by a hot gaseous halo that emerges at larger halo masses \citep{Keres:2005fk}.  These effects conspire to set the efficiency to about 20\% over this mass range.
(2) \citet{Dekel:2006fk} found that at high redshifts ($z>2$), narrow cold streams could penetrate directly into the halo, even when the halo mass is above the critical shock-heating mass $M_{\rm shock}\sim10^{12}$  \citep{Keres:2005fk},
therefore, the baryonic accretion efficiency should not change significantly across the critical shock-heating mass,
which is in agreement with our relatively constant cosmic star formation efficiency as a function of halo mass. In this case,
galaxies with a wide range of luminosities ($0.1L^*-3L^*$) should be dominated by a similar feedback process.


The predicted cosmic star formation efficiency from cosmological simulations
\citep[e.g.,][]{Oppenheimer:2008lr} is about a factor of two higher than 
the measurements. Therefore, there is
probably stronger feedback due to outflows in these galaxies, or other feedback effects 
need to be introduced to regulate the star formation process. 
We also note that the systematic uncertainty of the SFR measurements,  
e.g., different IMFs and dust extinction, can also offset the measured 
cosmic star formation efficiency by a factor of 2 or even larger \citep[e.g.,][]{Narayanan:2012lr}.
It is worth noting that although the cosmic star formation efficiency does not change significantly
with halo mass, the relation between UV luminosity and halo mass ($L\propto M^{1.5}$)
in LBGs at $z\sim3$, $z\sim4$, and $z\sim5$ implies that the cosmic star formation efficiency weakly
increases with halo mass. 

In Figure~\ref{fig:coldflow}, we also show the redshift-scaled SFR
and the host dark matter halo mass in submillimeter galaxies \citep[SMGs,][]{Hickox:2012ly}.
The redshift-scaled SFR in SMGs is about 0.85~dex higher than 
that in LBGs, suggesting that the intense starburst in the SMGs is 
not mainly fueled by cold accretion gas but triggered by the 
major merger process.
There are two different star formation modes found, which are long-lasting
modes in disk
galaxies (e.g., BzK galaxies) fueled by cold flows, and starburst modes in merging galaxies (e.g., SMGs)
triggered by major mergers \citep[e.g.][]{Daddi:2010lr}, and 
the SFRs in starbursts are 4-10 times larger than those in disk galaxies at fixed molecular gas mass.
Therefore, the plot demonstrates that LBGs with $L<2.5L^{*}$
follow a long-lasting mode in disk galaxies, more similar to local spirals and
BzK galaxies.

\section{SUMMARY}\label{summarysection}

We have carried out an LBC imaging survey
in the NDWFS Bo\"otes field, covering the 9 deg$^2$ field
with $U_{\rm spec}$ and $Y$-band, to fill in the two
main optical wavelength gaps in the NDWFS. The $5\sigma$ depth of 
$U_{\rm spec}$ is 25.5 mag.
The wide field allows us to 
select a statistically significant sample of luminous LBGs.
Using this sample, we have studied the bright end LF and clustering properties
of LBGs. The main scientific results are summarized as follows:
\begin{enumerate}
\item Using the LBT $U_{\rm spec}$ band images and the
NDWFS $B{\rm w}$ and $R$ images, we selected $14,485$
LBGs at redshift of $\sim3$ in the
9 deg$^2$ NDWFS Bo\"otes field, which
is the photometrically-selected LBG sample at $z\sim3$ in the largest area to date.

\item Combined with the faint-end LF measurements from \citet{Reddy:2009kx},
we measured the rest-frame UV LF of $z\sim3$ LBGs.
Our large field survey puts a strong constraint on the bright end 
of the LF. The Schechter function is fit to the UV LF, and the best-fit
parameters are $\Phi=(1.06\pm0.33)\times10^{-3}$~Mpc$^{-3}$, $M^{\star} = -20.11\pm0.08$, and $\alpha = -1.94\pm0.10$
by fitting our bright end data alone, and 
$\Phi=(1.12\pm0.17)\times10^{-3}$~Mpc$^{-3}$, $M^{\star} = -20.08\pm0.05$, and $\alpha = -1.90\pm0.05$
by combining our bright end data and the faint end data points from \citet{Reddy:2009kx}.
Both results support a steep faint-end slope of the LBG UV LF, rather than 
a relatively shallower faint-end slope as suggested by \citet{Shim:2007uq} and \citet{Sawicki:2006fj,Sawicki:2006lr}.
This implies that the faint LBGs make a significant contribution to the UV LF and 
dominate the SFR density at $z\sim3$.
With the large survey area and sample of LBGs, this work gives accurate measurement of 
the bright-end UV LF of $z\sim3$ LBGs, and allows us to probe the LBG luminosity range 
$-23.0<M_{\rm UV}<-25.0$.
At the brightest end,  the AGN/Quasar population dominates the LF. 
After subtracting the quasar LF from our measured LF, we still found
an excess over the Schechter function.
Further spectroscopic observations of the galaxy candidates
will allow us to confirm the bright LBG LF and study whether
the LF follows the Schechter function at the bright end.


\item We estimated the rest-frame near-IR LF of the $z\sim3$ LBGs. The best-fit Schechter 
function parameters are $\phi^{\star}_J=(3.1\pm1.9)\times10^{-4}$~Mpc$^{-3}$, 
$M^{\star}_J = -24.00\pm0.30$, and $\alpha_J = -2.00\pm0.17$.
The near-IR LF shows 
significant evolution compared to that of the rest-frame near-IR of local galaxies.
We derived the SMF of the $z\sim3$ LBGs using the near-IR LF.
The density of LBGs is lower than that of $K$-selected galaxies at $z\sim3$
at the massive end (M$>10^{11}$~M$_{\sun}$), suggesting that UV-selected
star-forming galaxies do not make a dominant contribution to the SMF
of $z\sim3$ galaxies at the massive end. The LBG SMF
shows a steep slope ($\alpha=-1.65\pm0.09$) and dominates the galaxy
stellar mass density at the low mass end. By comparing our measured SMF of LBGs
with cosmological hydrodynamic simulations with a
momentum-driven wind model, we found that the 
SMF derived from the simulation is consistent with the measured SMF
of LBGs at the massive and intermediate mass range, but the simulation
overproduces galaxies at the low mass end and does not produce
enough massive red and dead galaxies.

\item We studied the evolution of the LBG UV LF and SMF with cosmic time.
We compared the evolution with an episodic galaxy
formation model and a smoothy rising star formation model,
and found that the evolutionary trend supports the model with 
a steady rising star formation history. In this scenario, galaxies continuously
form stars in the redshift range between $z\sim7$ and $z\sim3$ and 
the SFR increases with increasing stellar mass to make the sSFR constant.

\item We also studied the clustering of two samples ($23.5<R<24$ and $24<R<24.5$)
of the bright LBGs. The comoving correlation lengths, $r_0$,
for these two samples are found to be $5.77\pm0.36 h^{-1}$~Mpc
and $5.14\pm0.16~h^{-1}$~Mpc, respectively. This result shows that
the galaxies with higher UV luminosity have stronger spatial clustering, implying
that the galaxies with higher UV luminosity are hosted by more massive dark matter halos.
We used HOD models to estimate the mean host dark matter halo
mass for these two LBG samples, and found that the mean host halo masses are $2.5\pm0.3\times10^{12}~h^{-1}M_{\sun}$ and 
$3.3^{+0.6}_{-0.4}\times10^{12}~h^{-1}M_{\sun}$, respectively. Combining with other 
clustering measurements of LBG samples at different redshifts, we found that 
the relationship of the  redshift-scaled SFR and the host halo mass
can be interpreted by cold flow gas accreted by the host dark matter halos, 
and the cosmic star formation efficiency in LBGs is about 5\%-20\% of the total cold flow gas.


\end{enumerate}






\acknowledgments
We thank the LBTO staff for their great support in preparing the 
observing and carrying out the observing with LBT/LBC.  
FB, XF, LJ and IM acknowledge support from a Packard Fellowship for Science and 
Engineering and NSF grant AST 08-06861 and AST 11-07682. 



{\it Facilities:} \facility{LBT}.



\appendix
\section{Integral Constraint}\label{ic}
Following the procedures in \citet{Adelberger:2005fk}, we calculate the total
IC from:
\begin{equation}
IC = \frac{1}{\sum_1^n RR_i}\sum_1^n\sigma_i^2DD_i,
\end{equation}
where the $RR_i$ is the random pair number in the $ith$ field
in the chosen angular bin, i.e., the sum sign means to sum over all
the individual field. The $\sigma_i$ for each field can be calculated
from, 
\begin{equation}
b = \frac{\sigma}{\sigma_{CDM}} = \frac{\sigma_{8,g}}{\sigma_8(z)},
\end{equation}
where $\sigma_8(z)$ is the linear matter
fluctuation in spheres of comoving $8~h^{-1}$~Mpc. 
We can get it from $\sigma_8(z)=\sigma_8(0)*D(z)$,
where $\sigma_8(0) = 0.9$, and $D(z)$ is linear growth factor.
$\sigma_{8,g}$ is the galaxies variance at the same scale, and
it is calculated from 
\begin{equation}
\sigma^2_{8,g} = \frac{72(r_0/8 h^{-1} Mpc)^\gamma}{(3-\gamma)(4-\gamma)(6-\gamma)2^{\gamma}},
\end{equation}
where $r_0$ and $\gamma$ are the comoving correlation length and the power-law index
in the 3-D correlation function.
$\sigma_{\rm CDM}$, the relative variance of the dark matter from one field to 
another, can be calculated from 
\begin{equation}
\sigma_{\rm CDM} = \frac{1}{(2\pi)^{3/2}}\left(\int d^3kP_L(|k|)|W_k(k)|\right)^{1/2},
\end{equation}
where $P_L(k)$ is the linear cold dark matter (CDM) power spectrum, and $W_k$
is the Fourier transform of a survey volume, which can be computed as
\begin{equation}
W_k = \exp\left(\frac{k_z^2l_z^2}{2}\right)\frac{sin(k_xI_x/2)}{k_xI_x/2}\frac{sin(k_yI_y/2)}{k_yI_y/2},
\end{equation}
where $I_x$ and $I_y$ are the comoving dimensions of the field of view for each field, and $I_z$ is comoving
dimensions of the radial direction, which can be convert from the redshift distribution. As the field of view
and selection function are almost the same for each field, we adopt the same $\sigma$ for all the fields.

\section{Halo Occupation Distribution}\label{hod}
The halo occupation distribution (HOD) is applied to our LBG
clustering results to interpret  host dark matter halo properties 
for these two bright LBG sub-samples. Following the procedure of \citet{Hamana:2004lr},
the dark matter halo mass function from \citet{Sheth:1999fk} is used.
The number distribution for a given galaxy population as a function of the host 
dark matter halo is adopted as:
\begin{eqnarray}
N_g(M) = \left \{ \begin{array}{lcl}
(M/M_1)^{\alpha} & \mbox{for}
& M\geqslant M_{\rm min} \\
0 & \mbox{for} & M<M_{\rm min}, \label{equation:hod}
\end{array}\right.
\end{eqnarray}
where $M_{\rm min}$ is the minimum mass of a halo
hosting the galaxy population, $M_1$ is the typical 
mass of a halo hosting one galaxy and $\alpha$ is the power-law
index.  
For close galaxy pairs in the
same dark matter halo, the following number distribution of the galaxy pairs
as a function of halo mass is applied \citep{Bullock:2002qy}:
\begin{eqnarray}
\langle N_g(N_g-1)\rangle(M) = \left \{ \begin{array}{lcl} 
N_g^2(M) ~~ \mbox{for} ~~ N_g(M) \geqslant 1 \\
N_g^2(M)\log(4N_g(M))/\log(4)\\ ~~~~~\mbox{for}~~1> N_g(M) \geqslant 0.25\\
0 ~~ \mbox{for} ~~ N_g(M) < 0.25.
\end{array}\right.
\end{eqnarray}

From the halo mass distribution described above and the galaxy population
distribution as a function of halo mass, we can further derive the 
number density of the galaxy population and the galaxy power spectrum,
which is comprised of two components, one is from the galaxy pairs in the 
same dark matter halo, the 1-halo term, and the other is from the galaxies
in two different dark matter halos, the 2-halo term. Then the galaxy spectrum is 
converted to the 2D ACF. We fit both the
2D ACF and number density of the galaxy population results
with our measurements.

\bibliography{paper}

\begin{thebibliography}{112}
\expandafter\ifx\csname natexlab\endcsname\relax\def\natexlab#1{#1}\fi

\bibitem[{{Adelberger} {et~al.}(1998){Adelberger}, {Steidel}, {Giavalisco},
  {Dickinson}, {Pettini}, \& {Kellogg}}]{Adelberger:1998qy}
{Adelberger}, K.~L., {Steidel}, C.~C., {Giavalisco}, M., {Dickinson}, M.,
  {Pettini}, M., \& {Kellogg}, M. 1998, \apj, 505, 18

\bibitem[{{Adelberger} {et~al.}(2005){Adelberger}, {Steidel}, {Pettini},
  {Shapley}, {Reddy}, \& {Erb}}]{Adelberger:2005fk}
{Adelberger}, K.~L., {Steidel}, C.~C., {Pettini}, M., {Shapley}, A.~E.,
  {Reddy}, N.~A., \& {Erb}, D.~K. 2005, \apj, 619, 697

\bibitem[{{Arnouts} {et~al.}(2005){Arnouts}, {Schiminovich}, {Ilbert},
  {Tresse}, {Milliard}, {Treyer}, {Bardelli}, {Budavari}, {Wyder}, {Zucca}, {Le
  F{\`e}vre}, {Martin}, {Vettolani}, {Adami}, {Arnaboldi}, {Barlow}, {Bianchi},
  {Bolzonella}, {Bottini}, {Byun}, {Cappi}, {Charlot}, {Contini}, {Donas},
  {Forster}, {Foucaud}, {Franzetti}, {Friedman}, {Garilli}, {Gavignaud},
  {Guzzo}, {Heckman}, {Hoopes}, {Iovino}, {Jelinsky}, {Le Brun}, {Lee},
  {Maccagni}, {Madore}, {Malina}, {Marano}, {Marinoni}, {McCracken}, {Mazure},
  {Meneux}, {Merighi}, {Morrissey}, {Neff}, {Paltani}, {Pell{\`o}}, {Picat},
  {Pollo}, {Pozzetti}, {Radovich}, {Rich}, {Scaramella}, {Scodeggio},
  {Seibert}, {Siegmund}, {Small}, {Szalay}, {Welsh}, {Xu}, {Zamorani}, \&
  {Zanichelli}}]{Arnouts:2005yq}
{Arnouts}, S., {et~al.} 2005, \apjl, 619, L43

\bibitem[{{Ashby} {et~al.}(2009){Ashby}, {Stern}, {Brodwin}, {Griffith},
  {Eisenhardt}, {Koz{\l}owski}, {Kochanek}, {Bock}, {Borys}, {Brand}, {Brown},
  {Cool}, {Cooray}, {Croft}, {Dey}, {Eisenstein}, {Gonzalez}, {Gorjian},
  {Grogin}, {Ivison}, {Jacob}, {Jannuzi}, {Mainzer}, {Moustakas},
  {R{\"o}ttgering}, {Seymour}, {Smith}, {Stanford}, {Stauffer}, {Sullivan},
  {van Breugel}, {Willner}, \& {Wright}}]{Ashby:2009kq}
{Ashby}, M.~L.~N., {et~al.} 2009, \apj, 701, 428

\bibitem[{{Becker} {et~al.}(1995){Becker}, {White}, \&
  {Helfand}}]{Becker:1995lr}
{Becker}, R.~H., {White}, R.~L., \& {Helfand}, D.~J. 1995, \apj, 450, 559

\bibitem[{{Bertin}(2006)}]{Bertin:2006qy}
{Bertin}, E. 2006, in Astronomical Society of the Pacific Conference Series,
  Vol. 351, Astronomical Data Analysis Software and Systems XV, ed.
  {C.~Gabriel, C.~Arviset, D.~Ponz, \& S.~Enrique}, 112

\bibitem[{{Bertin} \& {Arnouts}(1996)}]{Bertin:1996fk}
{Bertin}, E., \& {Arnouts}, S. 1996, \aaps, 117, 393

\bibitem[{{Bertin} {et~al.}(2002){Bertin}, {Mellier}, {Radovich}, {Missonnier},
  {Didelon}, \& {Morin}}]{Bertin:2002uq}
{Bertin}, E., {Mellier}, Y., {Radovich}, M., {Missonnier}, G., {Didelon}, P.,
  \& {Morin}, B. 2002, in Astronomical Society of the Pacific Conference
  Series, Vol. 281, Astronomical Data Analysis Software and Systems XI, ed.
  {D.~A.~Bohlender, D.~Durand, \& T.~H.~Handley}, 228

\bibitem[{{Bian} {et~al.}(2012){Bian}, {Fan}, {Jiang}, {Dey}, {Green},
  {Maiolino}, {Walter}, {McGreer}, {Wang}, \& {Lin}}]{Bian:2012lr}
{Bian}, F., {et~al.} 2012, \apj, 757, 139

\bibitem[{{Bielby} {et~al.}(2012){Bielby}, {Hill}, {Shanks}, {Crighton},
  {Infante}, {Bornancini}, {Francke}, {Heraudeau}, {Lambas}, {Metcalfe},
  {Minniti}, {Padilla}, {Theuns}, {Tummuangpak}, \&
  {Weilbacher}}]{Bielby:2012uq}
{Bielby}, R., {et~al.} 2012, ArXiv e-prints

\bibitem[{{Bielby} {et~al.}(2011){Bielby}, {Shanks}, {Weilbacher}, {Infante},
  {Crighton}, {Bornancini}, {Bouch{\'e}}, {H{\'e}raudeau}, {Lambas},
  {Lowenthal}, {Minniti}, {Padilla}, {Petitjean}, \& {Theuns}}]{Bielby:2011qy}
{Bielby}, R.~M., {et~al.} 2011, \mnras, 414, 2

\bibitem[{{Bouwens} {et~al.}(2007){Bouwens}, {Illingworth}, {Franx}, \&
  {Ford}}]{Bouwens:2007lr}
{Bouwens}, R.~J., {Illingworth}, G.~D., {Franx}, M., \& {Ford}, H. 2007, \apj,
  670, 928

\bibitem[{{Bouwens} {et~al.}(2008){Bouwens}, {Illingworth}, {Franx}, \&
  {Ford}}]{Bouwens:2008fk}
---. 2008, \apj, 686, 230

\bibitem[{{Bouwens} {et~al.}(2009){Bouwens}, {Illingworth}, {Franx}, {Chary},
  {Meurer}, {Conselice}, {Ford}, {Giavalisco}, \& {van
  Dokkum}}]{Bouwens:2009uq}
{Bouwens}, R.~J., {et~al.} 2009, \apj, 705, 936

\bibitem[{{Bouwens} {et~al.}(2011){Bouwens}, {Illingworth}, {Oesch},
  {Labb{\'e}}, {Trenti}, {van Dokkum}, {Franx}, {Stiavelli}, {Carollo},
  {Magee}, \& {Gonzalez}}]{Bouwens:2011uq}
---. 2011, \apj, 737, 90

\bibitem[{{Bouwens} {et~al.}(2012){Bouwens}, {Illingworth}, {Oesch}, {Franx},
  {Labb{\'e}}, {Trenti}, {van Dokkum}, {Carollo}, {Gonz{\'a}lez}, {Smit}, \&
  {Magee}}]{Bouwens:2012fk}
---. 2012, \apj, 754, 83

\bibitem[{{Brand} {et~al.}(2006){Brand}, {Brown}, {Dey}, {Jannuzi}, {Kochanek},
  {Kenter}, {Fabricant}, {Fazio}, {Forman}, {Green}, {Jones}, {McNamara},
  {Murray}, {Najita}, {Rieke}, {Shields}, \& {Vikhlinin}}]{Brand:2006ve}
{Brand}, K., {et~al.} 2006, \apj, 641, 140

\bibitem[{{Bruzual} \& {Charlot}(2003)}]{Bruzual:2003lr}
{Bruzual}, G., \& {Charlot}, S. 2003, \mnras, 344, 1000

\bibitem[{{Bullock} {et~al.}(2002){Bullock}, {Wechsler}, \&
  {Somerville}}]{Bullock:2002qy}
{Bullock}, J.~S., {Wechsler}, R.~H., \& {Somerville}, R.~S. 2002, \mnras, 329,
  246

\bibitem[{{Calzetti} {et~al.}(2000){Calzetti}, {Armus}, {Bohlin}, {Kinney},
  {Koornneef}, \& {Storchi-Bergmann}}]{Calzetti:2000vn}
{Calzetti}, D., {Armus}, L., {Bohlin}, R.~C., {Kinney}, A.~L., {Koornneef}, J.,
  \& {Storchi-Bergmann}, T. 2000, \apj, 533, 682

\bibitem[{{Cole} {et~al.}(2001){Cole}, {Norberg}, {Baugh}, {Frenk},
  {Bland-Hawthorn}, {Bridges}, {Cannon}, {Colless}, {Collins}, {Couch},
  {Cross}, {Dalton}, {De Propris}, {Driver}, {Efstathiou}, {Ellis},
  {Glazebrook}, {Jackson}, {Lahav}, {Lewis}, {Lumsden}, {Maddox}, {Madgwick},
  {Peacock}, {Peterson}, {Sutherland}, \& {Taylor}}]{Cole:2001fk}
{Cole}, S., {et~al.} 2001, \mnras, 326, 255

\bibitem[{{Cool}(2006)}]{Cool:2006bh}
{Cool}, R.~J. 2006, in Bulletin of the American Astronomical Society, Vol.~38,
  American Astronomical Society Meeting Abstracts, 1170

\bibitem[{{Cool}(2007)}]{Cool:2007zr}
{Cool}, R.~J. 2007, \apjs, 169, 21

\bibitem[{{Daddi} {et~al.}(2007){Daddi}, {Dickinson}, {Morrison}, {Chary},
  {Cimatti}, {Elbaz}, {Frayer}, {Renzini}, {Pope}, {Alexander}, {Bauer},
  {Giavalisco}, {Huynh}, {Kurk}, \& {Mignoli}}]{Daddi:2007qy}
{Daddi}, E., {et~al.} 2007, \apj, 670, 156

\bibitem[{{Daddi} {et~al.}(2010){Daddi}, {Elbaz}, {Walter}, {Bournaud},
  {Salmi}, {Carilli}, {Dannerbauer}, {Dickinson}, {Monaco}, \&
  {Riechers}}]{Daddi:2010lr}
---. 2010, \apjl, 714, L118

\bibitem[{{Dav{\'e}} {et~al.}(2011){Dav{\'e}}, {Oppenheimer}, \&
  {Finlator}}]{Dave:2011fj}
{Dav{\'e}}, R., {Oppenheimer}, B.~D., \& {Finlator}, K. 2011, \mnras, 415, 11

\bibitem[{{de Vries} {et~al.}(2002){de Vries}, {Morganti}, {R{\"o}ttgering},
  {Vermeulen}, {van Breugel}, {Rengelink}, \& {Jarvis}}]{de-Vries:2002lr}
{de Vries}, W.~H., {Morganti}, R., {R{\"o}ttgering}, H.~J.~A., {Vermeulen}, R.,
  {van Breugel}, W., {Rengelink}, R., \& {Jarvis}, M.~J. 2002, \aj, 123, 1784

\bibitem[{{Dekel} \& {Birnboim}(2006)}]{Dekel:2006fk}
{Dekel}, A., \& {Birnboim}, Y. 2006, \mnras, 368, 2

\bibitem[{{Dekel} {et~al.}(2009){Dekel}, {Birnboim}, {Engel}, {Freundlich},
  {Goerdt}, {Mumcuoglu}, {Neistein}, {Pichon}, {Teyssier}, \&
  {Zinger}}]{Dekel:2009lr}
{Dekel}, A., {et~al.} 2009, \nat, 457, 451

\bibitem[{{Dickinson} {et~al.}(2003){Dickinson}, {Papovich}, {Ferguson}, \&
  {Budav{\'a}ri}}]{Dickinson:2003fk}
{Dickinson}, M., {Papovich}, C., {Ferguson}, H.~C., \& {Budav{\'a}ri}, T. 2003,
  \apj, 587, 25

\bibitem[{{Donley} {et~al.}(2008){Donley}, {Rieke}, {P{\'e}rez-Gonz{\'a}lez},
  \& {Barro}}]{Donley:2008fk}
{Donley}, J.~L., {Rieke}, G.~H., {P{\'e}rez-Gonz{\'a}lez}, P.~G., \& {Barro},
  G. 2008, \apj, 687, 111

\bibitem[{{Dunlop} {et~al.}(2013){Dunlop}, {Rogers}, {McLure}, {Ellis},
  {Robertson}, {Koekemoer}, {Dayal}, {Curtis-Lake}, {Wild}, {Charlot},
  {Bowler}, {Schenker}, {Ouchi}, {Ono}, {Cirasuolo}, {Furlanetto}, {Stark},
  {Targett}, \& {Schneider}}]{Dunlop:2013kx}
{Dunlop}, J.~S., {et~al.} 2013, \mnras, 432, 3520

\bibitem[{{Elston} {et~al.}(2006){Elston}, {Gonzalez}, {McKenzie}, {Brodwin},
  {Brown}, {Cardona}, {Dey}, {Dickinson}, {Eisenhardt}, {Jannuzi}, {Lin},
  {Mohr}, {Raines}, {Stanford}, \& {Stern}}]{Elston:2006mz}
{Elston}, R.~J., {et~al.} 2006, \apj, 639, 816

\bibitem[{{Fan} {et~al.}(2001){Fan}, {Strauss}, {Schneider}, {Gunn}, {Lupton},
  {Becker}, {Davis}, {Newman}, {Richards}, {White}, {Anderson}, {Annis},
  {Bahcall}, {Brunner}, {Csabai}, {Hennessy}, {Hindsley}, {Fukugita}, {Kunszt},
  {Ivezi{\'c}}, {Knapp}, {McKay}, {Munn}, {Pier}, {Szalay}, \&
  {York}}]{Fan:2001lr}
{Fan}, X., {et~al.} 2001, \aj, 121, 54

\bibitem[{{Ferguson} {et~al.}(2004){Ferguson}, {Dickinson}, {Giavalisco},
  {Kretchmer}, {Ravindranath}, {Idzi}, {Taylor}, {Conselice}, {Fall},
  {Gardner}, {Livio}, {Madau}, {Moustakas}, {Papovich}, {Somerville},
  {Spinrad}, \& {Stern}}]{Ferguson:2004lr}
{Ferguson}, H.~C., {et~al.} 2004, \apjl, 600, L107

\bibitem[{{Finlator} {et~al.}(2011){Finlator}, {Oppenheimer}, \&
  {Dav{\'e}}}]{Finlator:2011qy}
{Finlator}, K., {Oppenheimer}, B.~D., \& {Dav{\'e}}, R. 2011, \mnras, 410, 1703

\bibitem[{{Foucaud} {et~al.}(2003){Foucaud}, {McCracken}, {Le F{\`e}vre},
  {Arnouts}, {Brodwin}, {Lilly}, {Crampton}, \& {Mellier}}]{Foucaud:2003lr}
{Foucaud}, S., {McCracken}, H.~J., {Le F{\`e}vre}, O., {Arnouts}, S.,
  {Brodwin}, M., {Lilly}, S.~J., {Crampton}, D., \& {Mellier}, Y. 2003, \aap,
  409, 835

\bibitem[{{Franx} {et~al.}(2003){Franx}, {Labb{\'e}}, {Rudnick}, {van Dokkum},
  {Daddi}, {F{\"o}rster Schreiber}, {Moorwood}, {Rix}, {R{\"o}ttgering}, {van
  der Wel}, {van der Werf}, \& {van Starkenburg}}]{Franx:2003fk}
{Franx}, M., {et~al.} 2003, \apjl, 587, L79

\bibitem[{{Giallongo} {et~al.}(2008){Giallongo}, {Ragazzoni}, {Grazian},
  {Baruffolo}, {Beccari}, {de Santis}, {Diolaiti}, {di Paola}, {Farinato},
  {Fontana}, {Gallozzi}, {Gasparo}, {Gentile}, {Green}, {Hill}, {Kuhn},
  {Pasian}, {Pedichini}, {Radovich}, {Salinari}, {Smareglia}, {Speziali},
  {Testa}, {Thompson}, {Vernet}, \& {Wagner}}]{Giallongo:2008fk}
{Giallongo}, E., {et~al.} 2008, \aap, 482, 349

\bibitem[{{Giavalisco} {et~al.}(1998){Giavalisco}, {Steidel}, {Adelberger},
  {Dickinson}, {Pettini}, \& {Kellogg}}]{Giavalisco:1998uq}
{Giavalisco}, M., {Steidel}, C.~C., {Adelberger}, K.~L., {Dickinson}, M.~E.,
  {Pettini}, M., \& {Kellogg}, M. 1998, \apj, 503, 543

\bibitem[{{Gonzalez} {et~al.}(2010){Gonzalez}, {Brodwin}, {Brown}, {Dey},
  {Dickinson}, {Gettings}, {Stanford}, {Stern}, {Bock}, {Bussman}, {Cooray},
  {Eisenhardt}, {Jannuzi}, {Lin}, {Mainzer}, \& {Sullivan}}]{Gonzalez:2010lr}
{Gonzalez}, A.~H., {et~al.} 2010, in American Astronomical Society Meeting
  Abstracts, Vol. 216, American Astronomical Society Meeting Abstracts \#216,
  415.13

\bibitem[{{Gonz{\'a}lez} {et~al.}(2010){Gonz{\'a}lez}, {Labb{\'e}}, {Bouwens},
  {Illingworth}, {Franx}, {Kriek}, \& {Brammer}}]{Gonzalez:2010fk}
{Gonz{\'a}lez}, V., {Labb{\'e}}, I., {Bouwens}, R.~J., {Illingworth}, G.,
  {Franx}, M., {Kriek}, M., \& {Brammer}, G.~B. 2010, \apj, 713, 115

\bibitem[{{Hamana} {et~al.}(2004){Hamana}, {Ouchi}, {Shimasaku}, {Kayo}, \&
  {Suto}}]{Hamana:2004lr}
{Hamana}, T., {Ouchi}, M., {Shimasaku}, K., {Kayo}, I., \& {Suto}, Y. 2004,
  \mnras, 347, 813

\bibitem[{{Hickox} {et~al.}(2012){Hickox}, {Wardlow}, {Smail}, {Myers},
  {Alexander}, {Swinbank}, {Danielson}, {Stott}, {Chapman}, {Coppin}, {Dunlop},
  {Gawiser}, {Lutz}, {van der Werf}, \& {Wei{\ss}}}]{Hickox:2012ly}
{Hickox}, R.~C., {et~al.} 2012, \mnras, 421, 284

\bibitem[{{Hildebrandt} {et~al.}(2007){Hildebrandt}, {Pielorz}, {Erben},
  {Schneider}, {Eifler}, {Simon}, \& {Dietrich}}]{Hildebrandt:2007rt}
{Hildebrandt}, H., {Pielorz}, J., {Erben}, T., {Schneider}, P., {Eifler}, T.,
  {Simon}, P., \& {Dietrich}, J.~P. 2007, \aap, 462, 865

\bibitem[{{Hildebrandt} {et~al.}(2009){Hildebrandt}, {Pielorz}, {Erben}, {van
  Waerbeke}, {Simon}, \& {Capak}}]{Hildebrandt:2009qy}
{Hildebrandt}, H., {Pielorz}, J., {Erben}, T., {van Waerbeke}, L., {Simon}, P.,
  \& {Capak}, P. 2009, \aap, 498, 725

\bibitem[{{Hoopes} {et~al.}(2003){Hoopes}, {Heckman}, {Jannuzi}, {Dey},
  {Brown}, {Ford}, \& {GALEX Science Team}}]{Hoopes:2003qf}
{Hoopes}, C.~G., {Heckman}, T.~M., {Jannuzi}, B.~T., {Dey}, A., {Brown},
  M.~J.~I., {Ford}, A., \& {GALEX Science Team}. 2003, in Bulletin of the
  American Astronomical Society, Vol.~35, American Astronomical Society Meeting
  Abstracts, 1371

\bibitem[{{Hunt} {et~al.}(2004){Hunt}, {Steidel}, {Adelberger}, \&
  {Shapley}}]{Hunt:2004fk}
{Hunt}, M.~P., {Steidel}, C.~C., {Adelberger}, K.~L., \& {Shapley}, A.~E. 2004,
  \apj, 605, 625

\bibitem[{{Isobe} {et~al.}(1986){Isobe}, {Feigelson}, \&
  {Nelson}}]{Isobe:1986mz}
{Isobe}, T., {Feigelson}, E.~D., \& {Nelson}, P.~I. 1986, \apj, 306, 490

\bibitem[{{Jain} \& {Lima}(2011)}]{Jain:2011lr}
{Jain}, B., \& {Lima}, M. 2011, \mnras, 411, 2113

\bibitem[{{Jannuzi} \& {Dey}(1999)}]{Jannuzi:1999fr}
{Jannuzi}, B.~T., \& {Dey}, A. 1999, in Astronomical Society of the Pacific
  Conference Series, Vol. 191, Photometric Redshifts and the Detection of High
  Redshift Galaxies, ed. {R.~Weymann, L.~Storrie-Lombardi, M.~Sawicki, \&
  R.~Brunner}, 111

\bibitem[{{Kajisawa} {et~al.}(2009){Kajisawa}, {Ichikawa}, {Tanaka}, {Konishi},
  {Yamada}, {Akiyama}, {Suzuki}, {Tokoku}, {Uchimoto}, {Yoshikawa}, {Ouchi},
  {Iwata}, {Hamana}, \& {Onodera}}]{Kajisawa:2009lr}
{Kajisawa}, M., {et~al.} 2009, \apj, 702, 1393

\bibitem[{{Kenter} {et~al.}(2005){Kenter}, {Murray}, {Forman}, {Jones},
  {Green}, {Kochanek}, {Vikhlinin}, {Fabricant}, {Fazio}, {Brand}, {Brown},
  {Dey}, {Jannuzi}, {Najita}, {McNamara}, {Shields}, \&
  {Rieke}}]{Kenter:2005gf}
{Kenter}, A., {et~al.} 2005, \apjs, 161, 9

\bibitem[{{Kere{\v s}} {et~al.}(2005){Kere{\v s}}, {Katz}, {Weinberg}, \&
  {Dav{\'e}}}]{Keres:2005fk}
{Kere{\v s}}, D., {Katz}, N., {Weinberg}, D.~H., \& {Dav{\'e}}, R. 2005,
  \mnras, 363, 2

\bibitem[{{Kroupa}(2001)}]{Kroupa:2001vn}
{Kroupa}, P. 2001, \mnras, 322, 231

\bibitem[{{Lacey} {et~al.}(2011){Lacey}, {Baugh}, {Frenk}, \&
  {Benson}}]{Lacey:2011qy}
{Lacey}, C.~G., {Baugh}, C.~M., {Frenk}, C.~S., \& {Benson}, A.~J. 2011,
  \mnras, 412, 1828

\bibitem[{{Lacy} {et~al.}(2004){Lacy}, {Storrie-Lombardi}, {Sajina},
  {Appleton}, {Armus}, {Chapman}, {Choi}, {Fadda}, {Fang}, {Frayer},
  {Heinrichsen}, {Helou}, {Im}, {Marleau}, {Masci}, {Shupe}, {Soifer},
  {Surace}, {Teplitz}, {Wilson}, \& {Yan}}]{Lacy:2004fk}
{Lacy}, M., {et~al.} 2004, \apjs, 154, 166

\bibitem[{{Landy} \& {Szalay}(1993)}]{Landy:1993yq}
{Landy}, S.~D., \& {Szalay}, A.~S. 1993, \apj, 412, 64

\bibitem[{{Le F{\`e}vre} {et~al.}(2005){Le F{\`e}vre}, {Paltani}, {Arnouts},
  {Charlot}, {Foucaud}, {Ilbert}, {McCracken}, {Zamorani}, {Bottini},
  {Garilli}, {Le Brun}, {Maccagni}, {Picat}, {Scaramella}, {Scodeggio},
  {Tresse}, {Vettolani}, {Zanichelli}, {Adami}, {Bardelli}, {Bolzonella},
  {Cappi}, {Ciliegi}, {Contini}, {Franzetti}, {Gavignaud}, {Guzzo}, {Iovino},
  {Marano}, {Marinoni}, {Mazure}, {Meneux}, {Merighi}, {Pell{\`o}}, {Pollo},
  {Pozzetti}, {Radovich}, {Zucca}, {Arnaboldi}, {Bondi}, {Bongiorno},
  {Busarello}, {Gregorini}, {Lamareille}, {Mathez}, {Mellier}, {Merluzzi},
  {Ripepi}, \& {Rizzo}}]{Le-Fevre:2005ys}
{Le F{\`e}vre}, O., {et~al.} 2005, \nat, 437, 519

\bibitem[{{Lee} {et~al.}(2009){Lee}, {Giavalisco}, {Conroy}, {Wechsler},
  {Ferguson}, {Somerville}, {Dickinson}, \& {Urry}}]{Lee:2009qy}
{Lee}, K.-S., {Giavalisco}, M., {Conroy}, C., {Wechsler}, R.~H., {Ferguson},
  H.~C., {Somerville}, R.~S., {Dickinson}, M.~E., \& {Urry}, C.~M. 2009, \apj,
  695, 368

\bibitem[{{Lee} {et~al.}(2006){Lee}, {Giavalisco}, {Gnedin}, {Somerville},
  {Ferguson}, {Dickinson}, \& {Ouchi}}]{Lee:2006qy}
{Lee}, K.-S., {Giavalisco}, M., {Gnedin}, O.~Y., {Somerville}, R.~S.,
  {Ferguson}, H.~C., {Dickinson}, M., \& {Ouchi}, M. 2006, \apj, 642, 63

\bibitem[{{Lee} {et~al.}(2011){Lee}, {Dey}, {Reddy}, {Brown}, {Gonzalez},
  {Jannuzi}, {Cooper}, {Fan}, {Bian}, {Glikman}, {Stern}, {Brodwin}, \&
  {Cooray}}]{Lee:2011fj}
{Lee}, K.-S., {et~al.} 2011, \apj, 733, 99

\bibitem[{{Lee} {et~al.}(2012){Lee}, {Ferguson}, {Wiklind}, {Dahlen},
  {Dickinson}, {Giavalisco}, {Grogin}, {Papovich}, {Messias}, {Guo}, \&
  {Lin}}]{Lee:2012lr}
---. 2012, \apj, 752, 66

\bibitem[{{Limber}(1953)}]{Limber:1953fk}
{Limber}, D.~N. 1953, \apj, 117, 134

\bibitem[{{Ly} {et~al.}(2009){Ly}, {Malkan}, {Treu}, {Woo}, {Currie},
  {Hayashi}, {Kashikawa}, {Motohara}, {Shimasaku}, \& {Yoshida}}]{Ly:2009qy}
{Ly}, C., {et~al.} 2009, \apj, 697, 1410

\bibitem[{{Madau}(1995)}]{Madau:1995lr}
{Madau}, P. 1995, \apj, 441, 18

\bibitem[{{Madau} {et~al.}(1996){Madau}, {Ferguson}, {Dickinson}, {Giavalisco},
  {Steidel}, \& {Fruchter}}]{Madau:1996uq}
{Madau}, P., {Ferguson}, H.~C., {Dickinson}, M.~E., {Giavalisco}, M.,
  {Steidel}, C.~C., \& {Fruchter}, A. 1996, \mnras, 283, 1388

\bibitem[{{Marchesini} {et~al.}(2009){Marchesini}, {van Dokkum}, {F{\"o}rster
  Schreiber}, {Franx}, {Labb{\'e}}, \& {Wuyts}}]{Marchesini:2009kx}
{Marchesini}, D., {van Dokkum}, P.~G., {F{\"o}rster Schreiber}, N.~M., {Franx},
  M., {Labb{\'e}}, I., \& {Wuyts}, S. 2009, \apj, 701, 1765

\bibitem[{{Meurer} {et~al.}(1999){Meurer}, {Heckman}, \&
  {Calzetti}}]{Meurer:1999fk}
{Meurer}, G.~R., {Heckman}, T.~M., \& {Calzetti}, D. 1999, \apj, 521, 64

\bibitem[{{Mo} \& {White}(1996)}]{Mo:1996kx}
{Mo}, H.~J., \& {White}, S.~D.~M. 1996, \mnras, 282, 347

\bibitem[{{Murray} {et~al.}(2005{\natexlab{a}}){Murray}, {Quataert}, \&
  {Thompson}}]{Murray:2005lr}
{Murray}, N., {Quataert}, E., \& {Thompson}, T.~A. 2005{\natexlab{a}}, \apj,
  618, 569

\bibitem[{{Murray} {et~al.}(2005{\natexlab{b}}){Murray}, {Kenter}, {Forman},
  {Jones}, {Green}, {Kochanek}, {Vikhlinin}, {Fabricant}, {Fazio}, {Brand},
  {Brown}, {Dey}, {Jannuzi}, {Najita}, {McNamara}, {Shields}, \&
  {Rieke}}]{Murray:2005ly}
{Murray}, S.~S., {et~al.} 2005{\natexlab{b}}, \apjs, 161, 1

\bibitem[{{Narayanan} \& {Dav{\'e}}(2012)}]{Narayanan:2012lr}
{Narayanan}, D., \& {Dav{\'e}}, R. 2012, \mnras, 423, 3601

\bibitem[{{Neistein} \& {Dekel}(2008)}]{Neistein:2008lr}
{Neistein}, E., \& {Dekel}, A. 2008, \mnras, 388, 1792

\bibitem[{{Oke} \& {Gunn}(1983)}]{Oke:1983qy}
{Oke}, J.~B., \& {Gunn}, J.~E. 1983, \apj, 266, 713

\bibitem[{{Oppenheimer} \& {Dav{\'e}}(2008)}]{Oppenheimer:2008lr}
{Oppenheimer}, B.~D., \& {Dav{\'e}}, R. 2008, \mnras, 387, 577

\bibitem[{{Oppenheimer} {et~al.}(2010){Oppenheimer}, {Dav{\'e}}, {Kere{\v s}},
  {Fardal}, {Katz}, {Kollmeier}, \& {Weinberg}}]{Oppenheimer:2010fk}
{Oppenheimer}, B.~D., {Dav{\'e}}, R., {Kere{\v s}}, D., {Fardal}, M., {Katz},
  N., {Kollmeier}, J.~A., \& {Weinberg}, D.~H. 2010, \mnras, 406, 2325

\bibitem[{{Ouchi} {et~al.}(2005){Ouchi}, {Hamana}, {Shimasaku}, {Yamada},
  {Akiyama}, {Kashikawa}, {Yoshida}, {Aoki}, {Iye}, {Saito}, {Sasaki},
  {Simpson}, \& {Yoshida}}]{Ouchi:2005vn}
{Ouchi}, M., {et~al.} 2005, \apjl, 635, L117

\bibitem[{{Papovich} {et~al.}(2011){Papovich}, {Finkelstein}, {Ferguson},
  {Lotz}, \& {Giavalisco}}]{Papovich:2011uq}
{Papovich}, C., {Finkelstein}, S.~L., {Ferguson}, H.~C., {Lotz}, J.~M., \&
  {Giavalisco}, M. 2011, \mnras, 412, 1123

\bibitem[{{Peebles}(1980)}]{Peebles:1980kx}
{Peebles}, P.~J.~E. 1980, {The large-scale structure of the universe,
  (Princeton: Princeton Univ. Press) }

\bibitem[{{Poli} {et~al.}(2001){Poli}, {Menci}, {Giallongo}, {Fontana},
  {Cristiani}, \& {D'Odorico}}]{Poli:2001kx}
{Poli}, F., {Menci}, N., {Giallongo}, E., {Fontana}, A., {Cristiani}, S., \&
  {D'Odorico}, S. 2001, \apjl, 551, L45

\bibitem[{{Reddy} {et~al.}(2012){Reddy}, {Pettini}, {Steidel}, {Shapley},
  {Erb}, \& {Law}}]{Reddy:2012qy}
{Reddy}, N.~A., {Pettini}, M., {Steidel}, C.~C., {Shapley}, A.~E., {Erb},
  D.~K., \& {Law}, D.~R. 2012, \apj, 754, 25

\bibitem[{{Reddy} \& {Steidel}(2009)}]{Reddy:2009kx}
{Reddy}, N.~A., \& {Steidel}, C.~C. 2009, \apj, 692, 778

\bibitem[{{Reddy} {et~al.}(2008){Reddy}, {Steidel}, {Pettini}, {Adelberger},
  {Shapley}, {Erb}, \& {Dickinson}}]{Reddy:2008fj}
{Reddy}, N.~A., {Steidel}, C.~C., {Pettini}, M., {Adelberger}, K.~L.,
  {Shapley}, A.~E., {Erb}, D.~K., \& {Dickinson}, M. 2008, \apjs, 175, 48

\bibitem[{{Roche} \& {Eales}(1999)}]{Roche:1999lr}
{Roche}, N., \& {Eales}, S.~A. 1999, \mnras, 307, 703

\bibitem[{{Rodighiero} {et~al.}(2011){Rodighiero}, {Daddi}, {Baronchelli},
  {Cimatti}, {Renzini}, {Aussel}, {Popesso}, {Lutz}, {Andreani}, {Berta},
  {Cava}, {Elbaz}, {Feltre}, {Fontana}, {F{\"o}rster Schreiber},
  {Franceschini}, {Genzel}, {Grazian}, {Gruppioni}, {Ilbert}, {Le Floch},
  {Magdis}, {Magliocchetti}, {Magnelli}, {Maiolino}, {McCracken}, {Nordon},
  {Poglitsch}, {Santini}, {Pozzi}, {Riguccini}, {Tacconi}, {Wuyts}, \&
  {Zamorani}}]{Rodighiero:2011fk}
{Rodighiero}, G., {et~al.} 2011, \apjl, 739, L40

\bibitem[{{Salpeter}(1955)}]{Salpeter:1955kx}
{Salpeter}, E.~E. 1955, \apj, 121, 161

\bibitem[{{Sawicki} \& {Thompson}(2006{\natexlab{a}})}]{Sawicki:2006fj}
{Sawicki}, M., \& {Thompson}, D. 2006{\natexlab{a}}, \apj, 642, 653

\bibitem[{{Sawicki} \& {Thompson}(2006{\natexlab{b}})}]{Sawicki:2006lr}
---. 2006{\natexlab{b}}, \apj, 648, 299

\bibitem[{{Schmidt}(1968)}]{Schmidt:1968fj}
{Schmidt}, M. 1968, \apj, 151, 393

\bibitem[{{Schmitt}(1985)}]{Schmitt:1985zr}
{Schmitt}, J.~H.~M.~M. 1985, \apj, 293, 178

\bibitem[{{Scoville} {et~al.}(2007){Scoville}, {Aussel}, {Brusa}, {Capak},
  {Carollo}, {Elvis}, {Giavalisco}, {Guzzo}, {Hasinger}, {Impey}, {Kneib},
  {LeFevre}, {Lilly}, {Mobasher}, {Renzini}, {Rich}, {Sanders}, {Schinnerer},
  {Schminovich}, {Shopbell}, {Taniguchi}, \& {Tyson}}]{Scoville:2007fk}
{Scoville}, N., {et~al.} 2007, \apjs, 172, 1

\bibitem[{{Shapley} {et~al.}(2001){Shapley}, {Steidel}, {Adelberger},
  {Dickinson}, {Giavalisco}, \& {Pettini}}]{Shapley:2001lr}
{Shapley}, A.~E., {Steidel}, C.~C., {Adelberger}, K.~L., {Dickinson}, M.,
  {Giavalisco}, M., \& {Pettini}, M. 2001, \apj, 562, 95

\bibitem[{{Sheth} \& {Tormen}(1999)}]{Sheth:1999fk}
{Sheth}, R.~K., \& {Tormen}, G. 1999, \mnras, 308, 119

\bibitem[{{Shim} {et~al.}(2007){Shim}, {Im}, {Choi}, {Yan}, \&
  {Storrie-Lombardi}}]{Shim:2007uq}
{Shim}, H., {Im}, M., {Choi}, P., {Yan}, L., \& {Storrie-Lombardi}, L. 2007,
  \apj, 669, 749

\bibitem[{{Soifer} \& {Spitzer/NOAO Team}(2004)}]{Soifer:2004pd}
{Soifer}, B.~T., \& {Spitzer/NOAO Team}. 2004, in Bulletin of the American
  Astronomical Society, Vol.~36, American Astronomical Society Meeting
  Abstracts \#204, 746

\bibitem[{{Spergel} {et~al.}(2007){Spergel}, {Bean}, {Dor{\'e}}, {Nolta},
  {Bennett}, {Dunkley}, {Hinshaw}, {Jarosik}, {Komatsu}, {Page}, {Peiris},
  {Verde}, {Halpern}, {Hill}, {Kogut}, {Limon}, {Meyer}, {Odegard}, {Tucker},
  {Weiland}, {Wollack}, \& {Wright}}]{Spergel:2007fk}
{Spergel}, D.~N., {et~al.} 2007, \apjs, 170, 377

\bibitem[{{Stark} {et~al.}(2009){Stark}, {Ellis}, {Bunker}, {Bundy}, {Targett},
  {Benson}, \& {Lacy}}]{Stark:2009lr}
{Stark}, D.~P., {Ellis}, R.~S., {Bunker}, A., {Bundy}, K., {Targett}, T.,
  {Benson}, A., \& {Lacy}, M. 2009, \apj, 697, 1493

\bibitem[{{Stark} {et~al.}(2007){Stark}, {Loeb}, \& {Ellis}}]{Stark:2007fj}
{Stark}, D.~P., {Loeb}, A., \& {Ellis}, R.~S. 2007, \apj, 668, 627

\bibitem[{{Steidel} {et~al.}(1999){Steidel}, {Adelberger}, {Giavalisco},
  {Dickinson}, \& {Pettini}}]{Steidel:1999lr}
{Steidel}, C.~C., {Adelberger}, K.~L., {Giavalisco}, M., {Dickinson}, M., \&
  {Pettini}, M. 1999, \apj, 519, 1

\bibitem[{{Steidel} {et~al.}(2003){Steidel}, {Adelberger}, {Shapley},
  {Pettini}, {Dickinson}, \& {Giavalisco}}]{Steidel:2003kx}
{Steidel}, C.~C., {Adelberger}, K.~L., {Shapley}, A.~E., {Pettini}, M.,
  {Dickinson}, M., \& {Giavalisco}, M. 2003, \apj, 592, 728

\bibitem[{{Steidel} {et~al.}(1996{\natexlab{a}}){Steidel}, {Giavalisco},
  {Dickinson}, \& {Adelberger}}]{Steidel:1996fk}
{Steidel}, C.~C., {Giavalisco}, M., {Dickinson}, M., \& {Adelberger}, K.~L.
  1996{\natexlab{a}}, \aj, 112, 352

\bibitem[{{Steidel} {et~al.}(1996{\natexlab{b}}){Steidel}, {Giavalisco},
  {Pettini}, {Dickinson}, \& {Adelberger}}]{Steidel:1996lr}
{Steidel}, C.~C., {Giavalisco}, M., {Pettini}, M., {Dickinson}, M., \&
  {Adelberger}, K.~L. 1996{\natexlab{b}}, \apjl, 462, L17+

\bibitem[{{Steidel} {et~al.}(2004){Steidel}, {Shapley}, {Pettini},
  {Adelberger}, {Erb}, {Reddy}, \& {Hunt}}]{Steidel:2004fk}
{Steidel}, C.~C., {Shapley}, A.~E., {Pettini}, M., {Adelberger}, K.~L., {Erb},
  D.~K., {Reddy}, N.~A., \& {Hunt}, M.~P. 2004, \apj, 604, 534

\bibitem[{{Stern} {et~al.}(2005){Stern}, {Eisenhardt}, {Gorjian}, {Kochanek},
  {Caldwell}, {Eisenstein}, {Brodwin}, {Brown}, {Cool}, {Dey}, {Green},
  {Jannuzi}, {Murray}, {Pahre}, \& {Willner}}]{Stern:2005lr}
{Stern}, D., {et~al.} 2005, \apj, 631, 163

\bibitem[{{Trenti} \& {Stiavelli}(2008)}]{Trenti:2008lr}
{Trenti}, M., \& {Stiavelli}, M. 2008, \apj, 676, 767

\bibitem[{{van der Burg} {et~al.}(2010){van der Burg}, {Hildebrandt}, \&
  {Erben}}]{van-der-Burg:2010vn}
{van der Burg}, R.~F.~J., {Hildebrandt}, H., \& {Erben}, T. 2010, \aap, 523,
  A74

\bibitem[{{van Dokkum}(2001)}]{van-Dokkum:2001lr}
{van Dokkum}, P.~G. 2001, \pasp, 113, 1420

\bibitem[{{van Dokkum} {et~al.}(2006){van Dokkum}, {Quadri}, {Marchesini},
  {Rudnick}, {Franx}, {Gawiser}, {Herrera}, {Wuyts}, {Lira}, {Labb{\'e}},
  {Maza}, {Illingworth}, {F{\"o}rster Schreiber}, {Kriek}, {Rix}, {Taylor},
  {Toft}, {Webb}, \& {Yi}}]{van-Dokkum:2006lr}
{van Dokkum}, P.~G., {et~al.} 2006, \apjl, 638, L59

\bibitem[{{Yan} {et~al.}(2011){Yan}, {Yan}, {Zamojski}, {Windhorst},
  {McCarthy}, {Fan}, {R{\"o}ttgering}, {Koekemoer}, {Robertson}, {Dav{\'e}}, \&
  {Cai}}]{Yan:2011qy}
{Yan}, H., {et~al.} 2011, \apjl, 728, L22

\bibitem[{{Yan} {et~al.}(2012){Yan}, {Finkelstein}, {Huang}, {Ryan},
  {Ferguson}, {Koekemoer}, {Grogin}, {Dickinson}, {Newman}, {Somerville},
  {Dav{\'e}}, {Faber}, {Papovich}, {Guo}, {Giavalisco}, {Lee}, {Reddy},
  {Cooray}, {Siana}, {Hathi}, {Fazio}, {Ashby}, {Weiner}, {Lucas}, {Dekel},
  {Pentericci}, {Conselice}, {Kocevski}, \& {Lai}}]{Yan:2012oq}
---. 2012, \apj, 761, 177

\bibitem[{{Zheng} {et~al.}(2007){Zheng}, {Coil}, \& {Zehavi}}]{Zheng:2007lr}
{Zheng}, Z., {Coil}, A.~L., \& {Zehavi}, I. 2007, \apj, 667, 760

\end{thebibliography}

\begin{table*}
\begin{center}
\caption{Summary of the $U_{\rm spec}$, $B{\rm w}$ and $R$-band $5\sigma$ depth\label{depth}}
\begin{tabular}{lcccc}
\tableline\tableline
Band &$\lambda_0$\tablenotemark{a} Median Image Quality      &Exposure time         & depth\tablenotemark{b}\\
&{\angstrom}&arcsec&second&\\
\tableline
$U_{\rm spec}$ &3590&1.25 & 1920 & 25.2 \\
$B{\rm w}$ &4133& 1.10 & 8400 & 26.3 \\
$R $ & 6407&1.10 & 6000 & 25.3 \\
\tableline
\end{tabular}
\tablenotetext{1}{Effective wavelength}
\tablenotetext{2}{Depth for $5\sigma$ detection in a $2\times$ FWHM aperture}
\end{center}
\end{table*}

\begin{table*}
\begin{center}
\caption{Schechter Parameters of UV LFs and Luminosity Densities}
\label{lftable}
\begin{tabular}{lcccccc}
\tableline\tableline
redshift & $\alpha$ & $M^{\star}(1700\rm\angstrom)$ & 
$\Phi^{\star}$ ($\times10^{-3}$)& $\rho_{L_{\rm{UV}}}$($\times10^{26}$) & ref\\
&&&$\rm{Mpc}^{-3}$&$\rm{erg}~s^{-1}~\rm{Hz}^{-1}~\rm{Mpc}^{-3}$&\\
\tableline
$2.7<z<3.3$ & $-1.94\pm0.10$ & $-21.11\pm0.08$ & $1.06\pm0.33$ & $2.19\pm0.08$& this work\\
$2.7<z<3.3$ & $-1.90\pm0.05$ &$-21.08\pm0.05$ & $1.12\pm0.17$& $2.18\pm0.05$ & this work + \citet{Reddy:2009kx}\\
$2.7<z<3.4$ & $-1.73\pm0.13$ & $-20.97\pm0.14$ & $1.71\pm0.53$ & $2.55\pm0.25$&\citet{Reddy:2009kx}\\
$3.0<z<3.4$ & $-0.83$ & $-20.69$ &$1.54$ &1.15 &\citealt{Shim:2007uq}\\
$2.7<z<3.3$ & $-1.43^{+0.17}_{-0.09}$&$-20.90^{+0.22}_{-0.14}$ &$1.70^{+0.59}_{-0.25}$&$1.81\pm0.04$&\citet{Sawicki:2006fj,Sawicki:2006lr}\\
\tableline
\end{tabular}
\end{center}
\end{table*}

\begin{table*}
\begin{center}
\caption{ACF and the Comoving Correlation Lengths for $z\sim3$ LBGs}\label{clustering}
\begin{tabular}{lcccccc}
\tableline
\tableline
brightness & $A_\omega$ & $\beta$ & $r_0$ & ref\\
\tableline
$23.5<R<24.0$ & $1.44\pm0.14$ & $0.6$ & $5.77\pm0.36$ & this work\\
$24.0<R<24.5$ & $1.13\pm0.06$ & $0.6$ & $5.14\pm0.16$ & this work\\
$R<24.0$ & $1.56^{+0.14}_{-0.17}$ & $0.6$ & $7.8^{+0.4}_{-0.5}$ & \citet{Lee:2006qy}\\
$R<24.5$ & $1.16^{+0.06}_{-0.08}$ & $0.6$ & $6.5^{+0.2}_{-0.3}$ &  \citet{Lee:2006qy}\\
$22.5<24.0$&-&$0.92\pm0.09$&$6.3\pm0.6$ & \citet{Hildebrandt:2007rt}\\
$22.5<24.5$&-&$0.55\pm0.08$&$5.2\pm0.4$ & \citet{Hildebrandt:2007rt}\\
\tableline
\end{tabular}
\end{center}
\end{table*}

\clearpage

\clearpage

\end{document}